\documentclass[conference]{IEEEtran}
\pdfoutput=1
\usepackage{trackchanges}
\addeditor{Jenny}
\addeditor{Zahra}

\ifCLASSINFOpdf
\else

\fi

\newcommand{\PRLsep}{\noindent\makebox[\linewidth]{\resizebox{0.3333\linewidth}{1pt}{$\bullet$}}\bigskip}

\usepackage{booktabs}
\usepackage{enumitem}
\usepackage{varwidth}
\usepackage{capt-of}
\usepackage{listings}
\usepackage{amsmath}
\usepackage{framed}
\usepackage{tabu}
\usepackage{float}
\usepackage[caption = false]{subfig}
\usepackage{enumitem}
\usepackage{color}
\usepackage{graphicx}
\usepackage{cleveref}
\usepackage{fancybox}
\usepackage[export]{adjustbox}
\usepackage[table]{xcolor}
\definecolor{Gray}{gray}{0.9}
\definecolor{White}{RGB}{255,255,255}
\usepackage{bbold}
\usepackage{bbm}
\usepackage{mathbbol}
\usepackage{multirow}
\usepackage{fix-cm}
\usepackage{siunitx,booktabs}
\usepackage{xcolor}

\usepackage{setspace}
\usepackage{graphicx}
\usepackage{colortbl}
\usepackage{array}
\usepackage{pifont}
\usepackage {rotating}
\usepackage{mwe}
\usepackage{pbox}
\usepackage{enumitem}
\let\oldding\ding% Store old \ding in \oldding
\renewcommand{\ding}[2][1]{\scalebox{#1}{\oldding{#2}}}
\usepackage{cite}

\newcommand{\Csharp}{%
  {\settoheight{\dimen0}{C}C\kern-.05em \resizebox{!}{\dimen0}{\raisebox{\depth}{\#}}}}

 \usepackage{enumitem}
\usepackage{xspace}
\usepackage{xparse}
\DeclareDocumentCommand\newstep{o}{%
\item\IfNoValueTF{#1}{}{#1 \textendash\xspace}}

\newlist{steps}{enumerate}{1}
\setlist[steps]{label=\textit{Step \arabic*:},leftmargin=*}

\definecolor{orange}{RGB}{0,32,96}
\definecolor{o}{RGB}{245,245,245}
\definecolor{g}{RGB}{50,50,50}

\PassOptionsToPackage{gray}{xcolor}
\usepackage{tikz}
\newcommand\mybox[2][]{\tikz[overlay]\node[fill=gray,inner sep=1pt, anchor=text, rectangle, rounded corners=0.5mm,#1] {#2};\phantom{#2}}

\usepackage{tcolorbox}

\begin{document}
\title{Task Interruptions in Requirements Engineering: Reality versus Perceptions!}

\author{\IEEEauthorblockN{Zahra Shakeri Hossein Abad, Guenther Ruhe}
\IEEEauthorblockA{Department of Computer Science\\ University of Calgary, Calgary, Canada\\
Email: \{zshakeri, ruhe\}@ucalgary.ca}
\and
\IEEEauthorblockN{Mike Bauer}
\IEEEauthorblockA{Arcurve Inc.\\
Calgary, Canada\\
Email: mike.bauer@arcurve.com}\vspace{-1ex}}


%Telephone: (800) 555--1212, Fax: (888) 555--1212}


\maketitle

\begin{abstract}
Task switching and interruptions are a daily reality in software development projects: developers switch between Requirements Engineering (RE), coding, testing, daily meetings, and other tasks. Task switching may increase productivity through increased information flow and effective time management. However, it might also cause a cognitive load to reorient the primary task, which accounts for the decrease in developers' productivity and increases in errors. This cognitive load is even greater in cases of cognitively demanding tasks as the ones typical for RE activities.
In this paper, to compare the reality of task switching in RE with the perception of developers, we conducted two studies: (i) a case study analysis on 5,076 recorded tasks of 19 developers and (ii) a survey of 25 developers. The results of our retrospective analysis show that in ALL of the cases that the disruptiveness of RE interruptions is statistically different from other software development tasks, RE related tasks are more vulnerable to interruptions compared to other task types. Moreover, we found that context switching, the priority of the interrupting task, and the interruption source and timing are key factors that impact RE interruptions. We also provided a set of RE task switching patterns along with recommendations for both practitioners and researchers.
While the results of our retrospective analysis show that self-interruptions are more disruptive than external interruptions, developers have different perceptions about the disruptiveness of various sources of interruptions. 
%Task interruptions are inevitable in every software development project. In recent years, researchers from variety of disciplines such as human computer interaction, cognitive psychology, and marketing have focused on interruptions through both experimental and theoretical investigations. However, analysis of the potential disruptive characteristics of these interruptions and their impact on interrupted tasks in the area of software development projects have not yet been sufficiently studied. In this paper, we report on an exploratory analysis to identify the disruptive characteristics of interruptions and to understand the impact of these interruptions on primary tasks. Finally we discuss the environmental factors such as teams and tasks structures on these interruptions. Based on the results of this analysis, we propose a set of guidelines and recommendations for managing task interruptions in software development projects.

\end{abstract}



% \fi
\IEEEpeerreviewmaketitle

\section{Introduction}
More than ever, multitasking is thrust upon software engineers by commercial pressures, such as budgetary constraints; customer support models that value supporting old versions and therefore have ad-hoc resourcing demands; services-based business models that strive to reduce non-billable time; or human-resources situations such as illness of a key team member or vacant team positions. A task switching may be caused by a change in task priority, a question from a coworker, a scheduled meeting, or a task blockage resulting from an unavailable resource needed for development \cite{Resumption}. Regardless of the source, task switching imposes a cognitive load and can be detrimental to the primary task, particularly in cases where at least one cognitively demanding task is involved in the task switching process. However, this does not mean that task switching is a bad thing. To the contrary, sequential multitasking often allows us to perform tasks effectively and it may foster efficiency in certain circumstances \cite{Mind} (e.g. knowledge transfer, managing the blocked tasks).

%the effects are often the same: when resuming interrupted work, developers experience increased time to perform the task, increased errors, increased loss of knowledge, and increased forgetting to perform critical tasks .} 

In the entire software lifecycle, Requirements Engineering (RE) is possibly the most data and communication intensive area \cite{RO}, crossing many social and organizational boundaries. The high level of complexity and cognition in RE activities, along with the key role of this phase in the success of software development projects, raises this key research question for further investigation: {\it ``How does task switching impact RE activities?''}.  
%However, there are some conditions in which task switchings can pose negative impact on a project process. OÕConnaill and Froehlich  \cite{behind} found that 41\% of the time an interrupted task was not resumed right away, and only 55\% of the tasks were resumed on the same day.
Over the last decade, a considerable amount of research into the cognitive aspects of software engineering (e.g. Section \ref {sec:RW}) has been undertaken. For example, \cite{Olli, Parisa, ParisaZahra} aim to assist the RE process by reducing the cognitive load of requirements elicitation and communication. However, the body of research on RE is lacking in understanding the concepts and the disruptiveness of interruptions in RE activities. This understanding can help inform how requirements engineers' productivity can be impacted by task switchings and what conditions make these interruptions more disruptive. 

This paper reports on a mixed methods study of task switching and interruptions in the area of RE. We conducted a manual retrospective analysis on 5,076 recorded tasks of 19 developers to compare RE tasks with other software development tasks in terms of their vulnerability to interruptions; and to understand and explore how RE tasks are influenced by interruptions. 
Further, a survey of 25 professional software developers was conducted to understand developers' perceptions and reasons behind task switchings and interruptions. 

From both studies, we found that context switching, interruptions with a different priority, and afternoon interruptions make RE interruptions more disruptive.
This paper makes the following contributions: 

\begin{enumerate}
\item it presents the results of a manual retrospective analysis of \(5,076\) recorded tasks of 19 professional software developers, comparing  RE tasks with other task types in terms of their vulnerability to interruptions, and specifically investigating the disruptiveness of RE interruptions,
\vspace{1mm}
\item it presents the results of a survey of 25 professional software developers, comparing the perceptions of developers to the repository analysis results,
\vspace{1mm}
\item it synthesizes the results of both studies and provides a set of ``RE task switching'' patterns and practical recommendations for practitioners by which they might better manage their RE task switchings and interruptions.    
\end{enumerate}

% While our retrospective analysis shows the negative impact of self-interruptions, our survey participants stated that interruptions with external causes negatively impacts their primary task productivity. We also proposed a set of RE task switching patterns, which have resulted from the cross-factor analysis we conducted in this study. 

We provide background information about our study in Section \ref{sec:BG}, followed by our Research Method (Section \ref{sec:Methodology}). Our study design, including the data collection, preparation and analysis, is discussed in Section \ref{sec:Design}. We discuss the main contributions of our study and provide a set of recommendations related to each contribution in Section \ref{sec:Results}. Limitations of this study are discussed in Section \ref{sec:threats}, followed by conclusion and research agenda in Section \ref{sec:Conclusion}.

\section{Background}
\label{sec:BG}
In this section, we first describe concepts related to task switching and interruption. Then, we review the related work, which addressed interruption analysis in software engineering. 
%In this section, key concepts underpinning this paper are presented and discussed. 
%concepts related to task interruptions and review the previous works addressed interruption analysis in software development projects. 

\subsection{Terminology}

\label{sec:terminology}
 Interruptions are a form of task switching or sequential multitasking \cite{Mind}. While there is some disagreement in the literature, there appears to be a general agreement, with plenty of theoretical evidence \cite{memoryofgoals, Mind, Disruptive2} about the cognitive cost of task switching, which appears on individual productivity. This theoretical evidence describes the  \mybox[fill=gray!30]{disruptiveness} of task switching and interruption in terms of  the time cost and the cost they pose on developers' productivity.  
 Memory-for-goals \cite{memoryofgoals} is one of the major cognitive theories on interruptions, which explains switching to another task suspends the goal of the primary task and activates the secondary task goal. The time period between task switches is called \mybox[fill=gray!30]{interruption lag} \cite{Mind, Resumption, ResumptionLag} (\({\bf D_3}\) in Figure \ref{fig:Terminology}).  In empirical studies, interruption lag is measured as the transitional interval between when a subject stops working on the primary task and when they start the secondary task. The timing of an interruption in respect to the primary task is a key aspect in the study of task interruptions \cite{Mind, Resumption}. As discussed by Salvucci and Taatgen \cite{Mind}, a primary task can rehearse its problem state very briefly, usually for a few seconds. The longer the time between pausing and resuming the primary task, the more time required to reconstruct the problem state after the task has been resumed \cite{Disruptive2}.
In this study, we define the \mybox[fill=gray!30]{resumption lag} as the time (i.e. in day{\footnotesize (s)}) between interrupting and resuming the primary task (\({\bf D_2}\) in Figure \ref{fig:Terminology}). 

Complementing the theory of memory-for-goals, Salvucci and Taatgen \cite{Mind} used the concept of \mybox[fill=gray!30]{problem state} to explain the process of sequential task switching and the suspension and resumption processes. The {\it problem state} refers to the information required for performing a task, which can be utilized with only one task at a given time.  What the current task is -{\bf the goal}- and the information required for doing a task -{\bf the problem state}- are maintained in distinct areas of the brain. This implies that the human's cognition only allows multiple goals to be active in their brain, not multiple problem states. Thus, task switchings which need to utilize the problem state at the same time contribute to a cognitive cost. However, there are some tasks during every software development project which are entirely reactive, such as answering an email, or a phone call, or when a manager pays a visit. In addition, there are some tasks that might need to utilize the problem state resource but do not need to keep the information therein, such as daily stand-up meetings \cite{Mind}. Thus, in this paper, we only consider task switchings as \mybox[fill=gray!30]{interruptions} in which the suspended task needs to maintain the problem state. For instance, if a development task is interrupted by a meeting, we count it as an interruption but not for the reverse case.
 \begin{tcolorbox}[colback=white, title= {Dependent (\(D_{1-3}\)) and Independent Variables (\(v_{1-8}\))}]
\vspace{-2mm}
 \underline{\small \bf \textcolor{orange}{Dependent Variables}}
 
\begin{minipage}[t]{0.001cm}
    \vspace*{0pt}
    \end{minipage}\hfill%
    \begin{minipage}[t]{8cm}
\vspace{-3.3mm}
        \includegraphics[scale=.68]{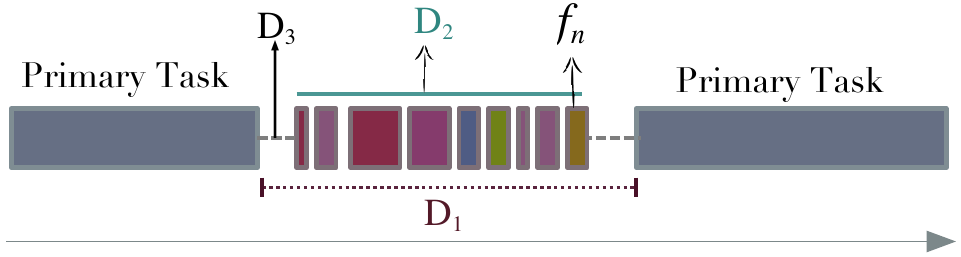}
        \vspace{-3mm}
     { \captionof{figure}{\noindent{ \({\bf D_1}\): \# of fragments \((=n)\), \({\bf D_2}\): resumption lag (in day(s)), \({\bf D_3}\): interruption lag, \({f_n}\): the \(n^{th}\) task fragment }}
 \label{fig:Terminology}}
    \end{minipage}\hfill
   
    \vspace{2mm}
    
    \underline{\small\bf  \textcolor{orange}{Independent Variables}}
        \vspace{1mm}
    \begin{spacing}{1}
{\footnotesize
{\bf Project variation (\(v_1\)): }{The {\it ProjectId} column from our database}
 \vspace{1mm}
 
{\bf Experience level (\(v_2\)): }\textcolor{g}{To measure the impact of the experience level, we analyzed the dataset of each employee from the last six months of their employment in the company. We recorded the experience level of each of the included employees in our study from their LinkedIn account. The average professional software development experience of participants is {\bf 3.5} (range 1 to 8) at Arcurve and {\bf 9.5} (range 4 to 25) in their total career.}
 \vspace{0.9mm}
 
{\bf Task level (\(v_3\)) \cite{Multitasking2, DecisionMaking, Concurrent}: }\textcolor{g}{The \(parentId\) column of our database}
 \vspace{0.9mm}

{\bf Interrupting task's type (\(v_4\)) \cite{Mind, Relation, Disruptive, Disruptive2}:} \textcolor{g}{We conducted a manual analysis on the meta data associated to each task to identify the type of both interrupted and interrupting tasks.}
 \vspace{0.9mm}
 
{\bf Interruption type (\(v_5\)) \cite{behind, Gonzalez, selfinterruption, Concurrent, Mind}:} \textcolor{g}{We used the meta data associated to each task. If the meta data did not help to record this variable, we queried the company's task management database and made the decision based on other instances of the primary and the interrupting task, before and after the interruption.}
 \vspace{0.9mm}

{\bf Interruption priority (\(v_6\)): }\textcolor{g}{The {\it Priority} column from our database}
 \vspace{-1.5mm}

{\bf daytime (\(v_7\)) \cite{behind}: }\textcolor{g}{All interruptions during the lunch time (i.e. 11:30 am- 1:00 pm) have been excluded from our recorded interruptions.}
 \vspace{1mm}

{\bf Task stage (\(v_8\)) \cite{Instant, Temporal}:}\textcolor{g} {{ To record this variable, we manually checked the time logs and textual meta data of each task. If the task was in formulation and clarification stages, we recorded them as {early stage interruptions}, otherwise as {late stage interruptions.}}}
 \vspace{-1.3mm}
 
 }
\end{spacing}
\end{tcolorbox}
\vspace{-1.5mm}
\noindent 

An interruption initiated by the subject of the primary task to address a more critical task or a planned task from the backlog is called a \mybox[fill=gray!30]{self-interruption}. 
Besides interrupting ourselves, an interruption can also be motivated by some external events in the environment (e.g. a colleague entering one's cubicle), which is called an \mybox[fill=gray!30]{external interruption} \cite{Gonzalez}. Many researchers \cite{behind, Concurrent, Gonzalez, Mind} used this classification as a key factor to measure the effect and the time cost of interruptions.

Considering the limited cognitive flexibility of humans \cite{ICSE}, nested task switchings causes mental congestion for keeping track of multiple states of tasks, which decays the goal of the primary task \cite{behind}. From existing interruption analysis studies and real-world examples \cite{Mind, behind}, it is clear that nested task switching causes a cognitive cost for reconstructing the problem state for every task switch which drives down the developers' productivity \cite{fragments}. In this study, in addition to ``interruption lag'' and ``resumption lag'', we use the \mybox[fill=gray!30]{number of task fragments} between pausing and resuming the primary task as a factor to measure the disruptiveness of interruptions (\({\bf D_1}\) in Figure \ref{fig:Terminology}).

\subsection{Related Work}
\label{sec:RW}
 As discussed in the background section above, there are plenty of both practical and theoretical evidence from Human Computer Interactions (HCI), Psychology, and Computer Supported Cooperative Work (CSCW) that studied the impact of task switching and interruption on individual productivity (e.g. \cite{Mind, behind, memoryofgoals, Disruptive2}). In addition to the studies discussed in the background section, in this section, we mainly focus on related work that studied task interruptions in the area of software engineering. Vasilescu et al. \cite{ICSE} modelled the rate and breadth of developers' context switching behaviour and studied the effects of task switchings on developers' productivity. They found that the {\it rate} and the {\it breadth} (i.e. the number of projects) of task switching is an influential factor on individuals' productivity: developers who are involved in several projects generate more output than who are not (breadth). However,  they found that frequent context switching during the course of a day has a negative impact on developers productivity (rate). They also surveyed developers to understand the main reasons for and perceptions of multitaskings in software development projects. Participants of this study describe the interrelationships and dependencies between projects as the most common reason for their task switchings. However, the results of this survey indicate that developers do not seem to be aware of the limits of multitasking (e.g. when to stop the task switching).

Meyer et al. \cite{Zimmer} conducted a survey and an observational study to understand developers' perception of productive and unproductive work. While their observational data shows that participants performed significant task switching during a day, yet, they perceive their days as productive when they complete many or big tasks without significant interruptions or context switches. Further, Parnin and Rugaber \cite{Resumption} conducted a survey and a retrospective analysis on 86 programmers to understand the various strategies and coping mechanisms that developers need to manage interrupted programming tasks. They found that only a small percentage of interrupted programming tasks were resumed in less than a minute. Also, among all resumption strategies they proposed, an automatic tag cloud of links to recent source code, contextual reminders, lightweight annotation of a task, and task snapshots are the top four choices of developers who participated in these studies. 

Chong and Siin \cite{Pair} compared interruption patterns among paired and solo programmers. Their study indicates significant differences between the pair programmers and solo programmers in terms of the length, type, time, context of occurrence and strategies for handling work interruptions. Also, they found that a substantial number of interruptions during the workday are self-initiated and joint work may have potential support for interruption handling.

While the existing research provided a wealth of insight on task switching and interruptions, we could not find any study that investigated interruptions in the area of RE. Moreover, the comprehensiveness of our study in terms of the size of our dataset and the number of dependent and independent variables makes it  different from other investigations.

\section{Research Method}
\label{sec:Methodology}
%The overall methodology of this research including the main steps and the artifacts of each step is depicted in Figure \ref{fig:process}. The details of each step (i.e. sub-steps and artifacts) are further discussed in this and following sections. During the process of implementing each of these steps, we considered {\it validity, feasibility,} and {\it usefulness} as the key requirements of every empirical study, as stated by Bickman and J. Rog \cite{Social}.

To design our research methodology, we followed the guidelines provided by Runeson and H\"{o}st \cite{Methodology}. Our research methodology comprises of both exploratory (i.e. retrospective analysis) and descriptive (i.e survey) approaches.  

\subsection{Context} 
We conducted this study in collaboration with Arcurve \footnote{\scriptsize www.arcurve.com/}, Calgary's largest independent software services company. This collaboration greatly helped us to achieve our study goals for several reasons: (1) Arcurve is specialized in custom software development, integration, implementation and managed application services, supporting the full software lifecycle; (2) the workers are involved in multiple projects related to diverse industries (e.g. oil and gas, mining, agriculture, asset management, health services, supply-chain logistics, and education; and (3) Arcurve's software development and delivery processes follow industry-leading practices, with an emphasis on iterative development, collaboration, and incremental delivery.

%
%
%
%\begin{figure*}
%\centering
%{\includegraphics[scale=1.1]{Process1}}
%\vspace{-1.5em}
%\caption{Main steps and the artifacts of the case study }
%\label{fig:process}
%\end{figure*} 

%\subsubsection{Defining Goals}
%We formulated the main goals of this research as follows:
%\begin{description}
%\item [Goal 1] {To conduct} a retrospective exploratory analysis of task interruptions to understand and characterize the behaviour of RE tasks {with respect to} interruption characteristics (e.g. frequency, resumption and interruption lag).
%\item [Goal 2] {To help} requirements engineers, practitioners, and researchers, {for} developing an intuitive understanding of task interruption {in} software development projects and to encourage more academic-industry collaboration. 
%
%\end{description}
\subsection{Developing the Conceptual Framework}
\label{sec:conceptual}
After developing the key concepts of our study, in this step, we developed the {\it conceptual framework}, including the key variables, factors and the potential relationships among them.

\subsubsection{Independent and Dependent Variables}
\label{sec:var}
 we identified a preliminary list of {dependent and independent variables} by conducting a comprehensive literature review on task switching and interruption analysis. These variables were pilot tested \cite{SERIP} on 7,770 recorded tasks of 10 employees to ensure the quality of our dataset and to identify the potential confounding variables, such as interruption source and type, experience level, and task stage.  The datasets required for both studies were collected from Arcurve's task-based bug tracking and project management tool (i.e. Fogbugz\footnote{https://shop.fogcreek.com/FogBugz/}). The three dependent {\small(\(D_{1-3}\))} and eight independent {\small(\(v_{1-8}\))} variables of this study, the way we interpreted them in the course of our data analysis, and their corresponding literature references are listed in Figure  \ref{fig:Terminology}.

%\textbf{\textit { Temporal Interrupt Position:}}
%to investigate the temporal interrupt positions of task interruption, Czerwinski et al. \cite{Instant} performed a series of experiments examining the temporal state of interruptions in terms of the interruption lag. They showed that when users were interrupted at the early stages of a task, they were faster at switching to the secondary task. Monk and Boehm-Davis \cite{Temporal} made similar observations to measure the time required to reconstruct/retrieve the problem state of interrupted tasks after their resumption in terms of the point at which a task is interrupted (i.e.  \mybox[fill=gray!30] {task stage}). They found interrupting a task during midel- or end-task interruption points will result in longer resumption time. Supplementing these studies, Mark et al. \cite{behind} looked at the frequency of interruptions that occur in the morning and afternoon and showed that there are not significant differences between numbers of interruptions that occur at different points in a day (i.e.  \mybox[fill=gray!30] {daytime}). In this study, we considered the {\it task stage} and {\it the time of day} to understand and measure the impact of interruptions. 

\subsubsection{Exploratory Factor Analysis}

due to the complexity of interpreting all combinations of our dependent and independent variables (i.e. {\small \(8\times3=24\)}), we used the {\bf Binary Exploratory Factor Analysis (BEFA)} approach. This method facilitates easier interpretation of the results by identifying the underlying dimensions of a dataset and classifying the variables based on these dimensions. To implement the BEFA approach, we manually analyzed \(4,093\) task logs (included in \(5,076\) task logs) of our dataset to explore 70 interruptions/task switchings per employee.  We chose a cut-off of 0.32 for a statistically meaningful rotated factor loading, following the rule of thumb proposed by Tabachnick et al. \cite{thumb}. Table \ref{tab:loading} presents the Varimax-rotated matrix between the factors and variables after the factors have been sorted by their absolute loading values. 
%Interpreting the factors based on the natural selection of variable loadings we could say that the conceptual meaning of factor 1 is related to {\it environment}, factor 2 is related to {\it interruption source and task features }, and factor 3 is related to the {\it temporal point} of interruptions.

Looking at the factor loadings of the variables in this table, we observe that {\bf Factor1} has high loadings, shaded in gray, on variables \((v_{1-3})\), which describes the {\it contextual} aspects of interruptions, such as other ongoing projects and tasks in the company, and the experience level of the employees. {\bf Factor2} has high loadings on variables \((v_{4-6})\) which explain {\bf interruption characteristics}, such as the type of the interrupting task, priority, and interruption source. Likewise, {\bf Factor3} has high loadings on variables \((v_3, v_{7-8})\) and describes the {\bf temporal} aspects of interruptions. To interpret and name this factor, we only used \(v_7\) and \(v_8\) as \(v_2\) is assigned to Factor1.

\begin {table}
\centering
\scriptsize
\vspace{-3mm}
\caption {Factor matrix showing factor loadings on the variables}
\label{tab:loading}
\vspace{-1.1em}
\begin{tabular} {|p{4.5cm}|p{1cm}|p{1cm}|p{1cm}|} \hline
{\textcolor{white}{..............}\bf Variables (possible values)}&{\bf Factor1} &{\bf Factor2}&{\bf Factor3} \\ \hline
Project variation (same/different)({\(v_1\)}) & \cellcolor{Gray} 0.50&0.27&-0.07\\
Employees' experience (less/more) ({\(v_2\)})  & \cellcolor{Gray}0.81&-0.13&-0.14\\
Task Hierarchy (main/{sub-task}) ({\(v_3\)}) & \cellcolor{Gray}-0.87&0.18&\cellcolor{Gray}-0.46 \\
Interruption type (self/{\bf external*}) ({\(v_4\)})  &0.09& \cellcolor{Gray}0.87&0.26\\
Task type (same/different) ({\(v_5\)})  & 0.07& \cellcolor{Gray}0.61&0.08 \\
Priority difference (same/different) ({\(v_6\)})  &0.25& \cellcolor{Gray}-0.38&-0.05 \\
Interruption time ({\bf morning}/afternoon) ({\(v_7\)})  &-0.10&-0.28& \cellcolor{Gray}0.39\\
Task Stage (early/late) ({\(v_8\)})  &0.01&0.02& \cellcolor{Gray}0.38\\\hline
\multicolumn{4}{p{7.7cm}}{\tiny \it*The bold values present the value of each variable that has been used for implementing the BEFA method}\\
\end{tabular}
\vspace{-6mm}
\end{table}

\subsection{Research Questions}
\label{sec:RQ}
Our study was guided with the following research questions:

{{\bf RQ1 (RE vs other tasks):} {Regarding the disruptiveness of interruptions, is there a significant difference between RE and other software development activities in terms of the context, type, and timing of interruptions?}}
\vspace{2mm}
 
% (1) quantitatively investigating the association between the explored factors in Section \ref{sec:FA} and the disruptiveness of task interruptions/switching, and (2) comparing the degree of this disruptiveness of task switchings/interruption in RE and other software development tasks.  

% 
% This RQ helps to examine how different types of interrupting tasks may have differential effects on how interruption impacts primary-task performance. On a related note, another aspect of task type that has been studied is the relevance of the interrupting task to the original (interrupted) task. \cite{Concurrent}
 
{\bf RQ2 (RE interruptions):} {How and to what extent RE interruptions are influenced by the context, type, and timing of interruptions?}
 
% RQ2 aims at (1) analyzing which type of variables make a significant impact on the disruptiveness of task switchings/interruptions in the area of RE, and (2) quantitatively investigating the impact of each of these variables. 
% 
\vspace{2mm}
  
{{\bf RQ3 (Cross-factor analysis):} {How do cross-factor combinations of variables \(v_{1-8}\) impact RE interruptions?}}

%In this se
%
% {Internal Validity:} In general, an analysis of validity assesses how well a research instrument identifies concepts of interest. Construct validity has implications for the measures used, findings and conclusions. Face, criterion and content validity relate to the measures employed. Internal and external validity provide support for an overall assessment of the results and conclusions. Validity protects the repeatability of the study findings (Babbie, 2001).
%
%- {\it Face validity:} \textcolor{red}{we achieved this level of validity though interviews software practitioners, a relevant literature review, and piloting the survey instrument. \cite{FaceValidity}}
%
%
%\subsubsection{External Validity:}
%

\begin {table*}
\scriptsize
\centering
\caption {RQ1- Comparison between RE and other software development tasks in terms of the disruptiveness of various factors}
\label{tab:RQ1}
\vspace{-2mm}
\begin{tabular} {|p{1.8cm}|p{0.7cm}p{0.65cm}p{0.6cm}|p{0.5cm}p{0.6cm}p{0.7cm}|p{0.6cm}p{0.6cm}p{0.5cm}|p{0.5cm}p{0.5cm}p{0.6cm}|p{0.5cm}p{0.6cm}p{0.6cm}|p{0.7cm}p{0.7cm}p{0.5cm}|p{0.8cm}p{0.65cm}p{0.5cm}|p{0.5cm}p{0.6cm}p{0.5cm}|} \hline
{}&\multicolumn{9}{c|}{{\bf Interruption Context}}&\multicolumn{9}{c|}{{\bf Interruption Type}}&\multicolumn{6}{c|}{{\bf Interruption Time}}\\\cline{2-25}

\textcolor{white}{.........}{\bf Pairs}&\multicolumn{3}{c|}{{different project}}&\multicolumn{3}{c|}{less experience}&\multicolumn{3}{c|}{sub-task}&\multicolumn{3}{c|}{{different type}}&\multicolumn{3}{c|}{{self-interruption}}&\multicolumn{3}{c|}{{different priority}}&\multicolumn{3}{c|}{{afternoon Int.}}&\multicolumn{3}{c|}{{late stage}}\\

{}&\multicolumn{3}{c|}{{(\(v_1 \)){\it }}}&\multicolumn{3}{c|}{{(\( v_2\))}}&\multicolumn{3}{c|}{{(\(v_3\))}}&\multicolumn{3}{c|}{{(\(v_4\))}}&\multicolumn{3}{c|}{{ (\(v_5\))}}&\multicolumn{3}{c|}{{ (\(v_6\))}}&\multicolumn{3}{c|}{{(\(v_7\))}}&\multicolumn{3}{c|}{{(\(v_8\))}}\\\cline{2-25}

{}&\(D_1\)&\(D_2\)&\(D_3\)&\(D_1\)&\(D_2\)&\(D_3\)&\(D_1\)&\(D_2\)&\(D_3\)&\(D_1\)&\(D_2\)&\(D_3\)&\(D_1\)&\(D_2\)&\(D_3\)&\(D_1\)&\(D_2\)&\(D_3\)&\(D_1\)&\(D_2\)&\(D_3\)&\(D_1\)&\(D_2\)&\(D_3\)\\\hline

{Kruskal-Wallis}&{0.003}&{0.01}&{0.32}&0.1&0.01&{0.01}&1e-4&4e-4&{0.03}&1e-3&3e-4&0.04&0.01&0.003&0.04&4e-6&3e-5&0.05&0.01&0.04&0.01&0.03&0.01&0.2\\\hline

{\bf RE-Architecture}&\cellcolor{Gray}0.04&0.02*&{0.3}&0.1&\cellcolor{Gray}0.02&{0.2}&\cellcolor{Gray}4e-5&\cellcolor{Gray}1e-4&{0.2}&\cellcolor{Gray}0.01&\cellcolor{Gray}0.01&0.8&0.17&0.1&{0.7}&0.05*&0.04*&{0.7}&\cellcolor{Gray}0.03&\cellcolor{Gray}0.02&{0.4}&0.05*&0.1&0.3\\

{\bf RE-Development}&0.1&0.1&{0.8}&0.2&0.1&\cellcolor{Gray}{0.001}&\cellcolor{Gray}7e-5&\cellcolor{Gray}0.001&{\cellcolor{Gray}0.01}&\cellcolor{Gray}0.01&\cellcolor{Gray}0.03&\cellcolor{Gray}{0.03}&0.2&0.2&{0.3}&\cellcolor{Gray}{5e-5}&\cellcolor{Gray}2e-5&{0.5}&{0.1}&0.05*&\cellcolor{Gray}{0.03}&0.5&0.3&0.3\\

{\bf RE-Test}&\cellcolor{Gray}4e-5&\cellcolor{Gray}3e-5&{0.3}&0.1&0.03*&{0.1}&\cellcolor{Gray}6e-5&\cellcolor{Gray}0.001&{0.1}&\cellcolor{Gray}4e-4&\cellcolor{Gray}1e-5&0.04*&0.01*&\cellcolor{Gray}0.01&{0.1}&\cellcolor{Gray}{0.002}&\cellcolor{Gray}0.002&{0.2}&{0.07}&{0.01*}&0.7&0.1&\cellcolor{Gray}{0.03}&0.6\\

{\bf RE-UI Design}&0.07&0.08&{0.3}&0.1&0.2&{0.1}&\cellcolor{Gray}1e-6&\cellcolor{Gray}2e-4&{0.4}&\cellcolor{Gray}0.01&\cellcolor{Gray}0.03&0.01*&0.1&0.1&{0.06}&\cellcolor{Gray}0.001&\cellcolor{Gray}0.003&{\cellcolor{Gray}0.05}&0.02*&0.01*&{0.06	}&0.06&0.06&0.1\\

{\bf RE-Deployment}&0.3&0.6&{0.01*}&0.17&0.2&\cellcolor{Gray}{0.04}&\cellcolor{Gray}0.01&\cellcolor{Gray}0.01&{0.02}&0.2&0.6&\cellcolor{Gray}0.01&0.8&0.7&{\cellcolor{Gray}0.05}&0.4&0.5&{0.1}&{0.7}&0.5&{0.06}&\cellcolor{Gray}0.03&0.3&0.1\\\hline
\multicolumn{25}{l}{*: {\it The p-value of the alternative value of the corresponding variable. For instance, if (\(v_2=\text{more exp}\)), there will be a significant difference between RE-Test in terms of \(D_2\) }}\\

\end{tabular}
\vspace{-5mm}
\end{table*}

\section{Study Design}
\label{sec:Design}
\subsection{Data Collection and Preparation} 
During the four months of data collection and preparation, we hired three Research Assistants (RA) in the area of software engineering and trained them over the first two months to familiarize them with the data preparation process. After the first two months, we ran a pilot study on the prepared dataset and reported the results \cite{SERIP} to our industry partner. Following the feedback we got from this step, we revised our data extraction form and started data extraction and preparation pahses from scratch. The data extraction form and a sample dataset collected for one employee are available on the website of the first author\footnote{\scriptsize http://wcm.ucalgary.ca/zshakeri/projects}. For each employee, we recorded 100 interruptions due to the high level of details in our data extraction form.

%{\small
%\underline{\it Data Collection Rules}
%\begin{itemize}
%
%\item Switching from management meetings, site visits, daily stand-up meetings were not considered as interruptions
%\item Weekends have not been counted in {\it resumption lag}
%\item 
%\end{itemize}}

%\begin {table}
%\scriptsize
%\centering
%\caption {Details of the manually analyzed tasks}
%\label{tab:dataset}
%\begin{tabular} {|p{1.8cm}|p{2.1cm}|p{2cm}|p{1.7cm}|} \hline
%{\bf Scope} & {\bf \# of analyzed tasks}  & {\bf \#of recorded tasks} & {\bf Subjects}\\ \hline
%{\bf Factor analysis} & \(\sim 4,000^*\)& \(1,330\)&\multirow{3}{*}{ Employees (19)} \\ 
%{\bf RQ1} &\multirow{2}{*}{ 5,076}&\multirow{2}{*}{ 1,900}& \\ 
%{\bf RQ2} &&& \\ \hline
%{\bf RQ3} & \(\sim6,000\)&  \(\sim 6,000\)& Project (4)\\ \hline
%{\bf Total} &\multicolumn{3}{c|}{\({\bf 11,000 \text \quad {tasks}}\)} \\ \hline
%\multicolumn{4}{p{7.5cm}}{\(^*\): the dataset used for the factor analysis step is a subset of the data used for RQ1-2} \\
%\end{tabular}
%
%\end{table}
%

\subsection{Data Analysis Methods} 
\label{sec:analysis}
To analyze and interpret the results of RQ1 and 2, we used {\bf statistical hypothesis testing}, where the null hypotheses for each of these questions were formulated as follows:
\begin{itemize}
 \item {\bf [RQ1]} \textit{ {\(H_0(1, v_i D_j, type_k)=\) \(<variable_i>\) does not make a significant difference between the \(<disruptiveness_j>\) resulted from task interruption of RE and \(<type_k>\) tasks.}} 
 
 \item {\bf[RQ2]} \textit{ {\(H_0(2, v_i D_j)=\) In the context of RE task interruptions, \(<variable_i>\) does not make a significant impact on \(<disruptiveness_j>\).}} 
  \end{itemize}
 
 Where \(i\in \{1, 2, ..., 8\}\) and \(j= 1, 2, 3\). Moreover, to answer RQ3, we used the Pearson cross-factor correlation analysis along with statistical tests. We used {\it Q-Q} plots \cite{qq} to assess the normality of the distribution of our dataset. As we could not confirm that the underlying populations follow a normal distribution, we used the {\it Kruskal-Wallis} test to examine the null hypotheses posed for RQ1-3. To properly address RQ1, if the overall statistically significant difference in group means was shown by the  {\it Kruskal-Wallis} test, we used the {\it Kruskal-Wallis post-hoc} test for pairwise multiple comparisons to examine each pair separately. To implement our statistical tests, we used the {\bf {\small PMCMR}}\footnote{\scriptsize https://cran.r-project.org/web/packages/PMCMR/PMCMR.pdf}package of {\bf R}.

\begin{figure}[!t]
\centering
\vspace{-2mm}
\subfloat[]{\includegraphics[scale=0.41]{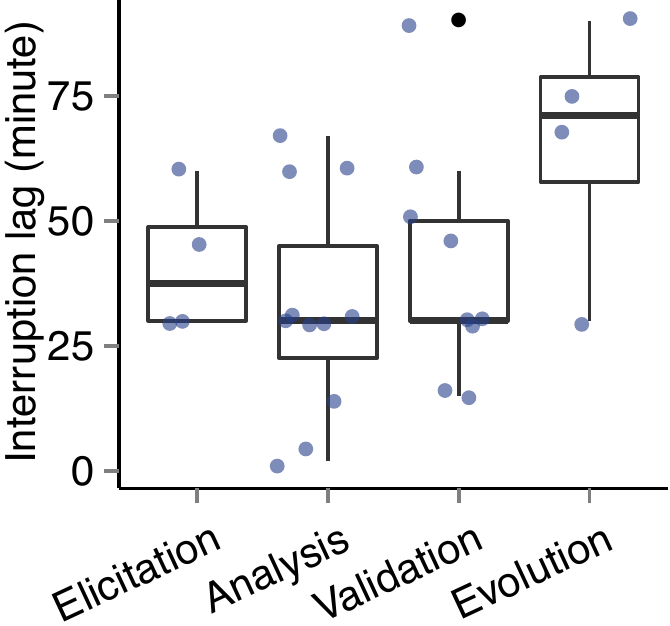}}
\subfloat[]{\includegraphics[scale=0.41]{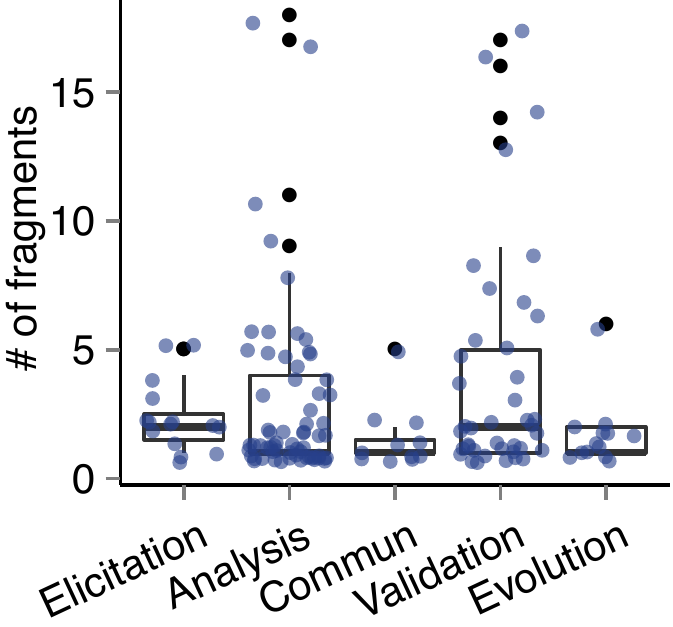}}
\subfloat[]{\includegraphics[scale=0.41]{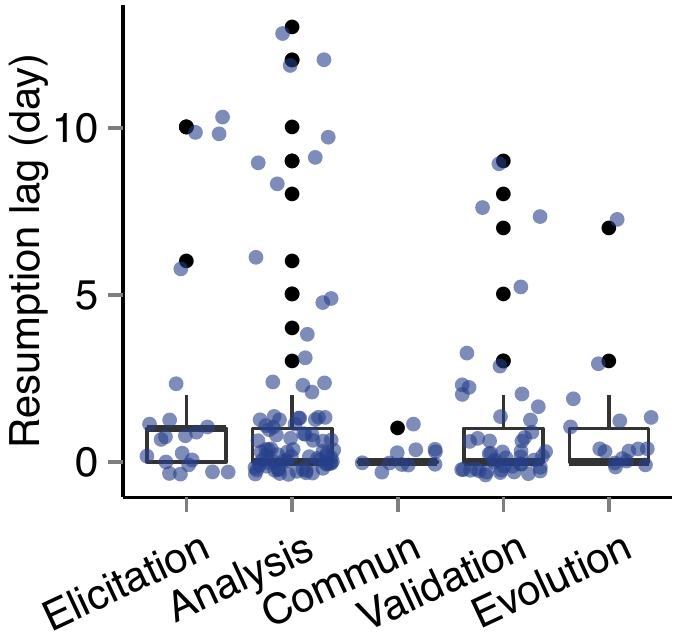}}
\vspace{-2mm}
\caption{Comparison between RE activities}
\label{fig:RETasks}
\vspace{-5mm}
\end{figure}

\begin{figure}
\vspace{-2mm}
\centering
\subfloat{\includegraphics[scale=0.13]{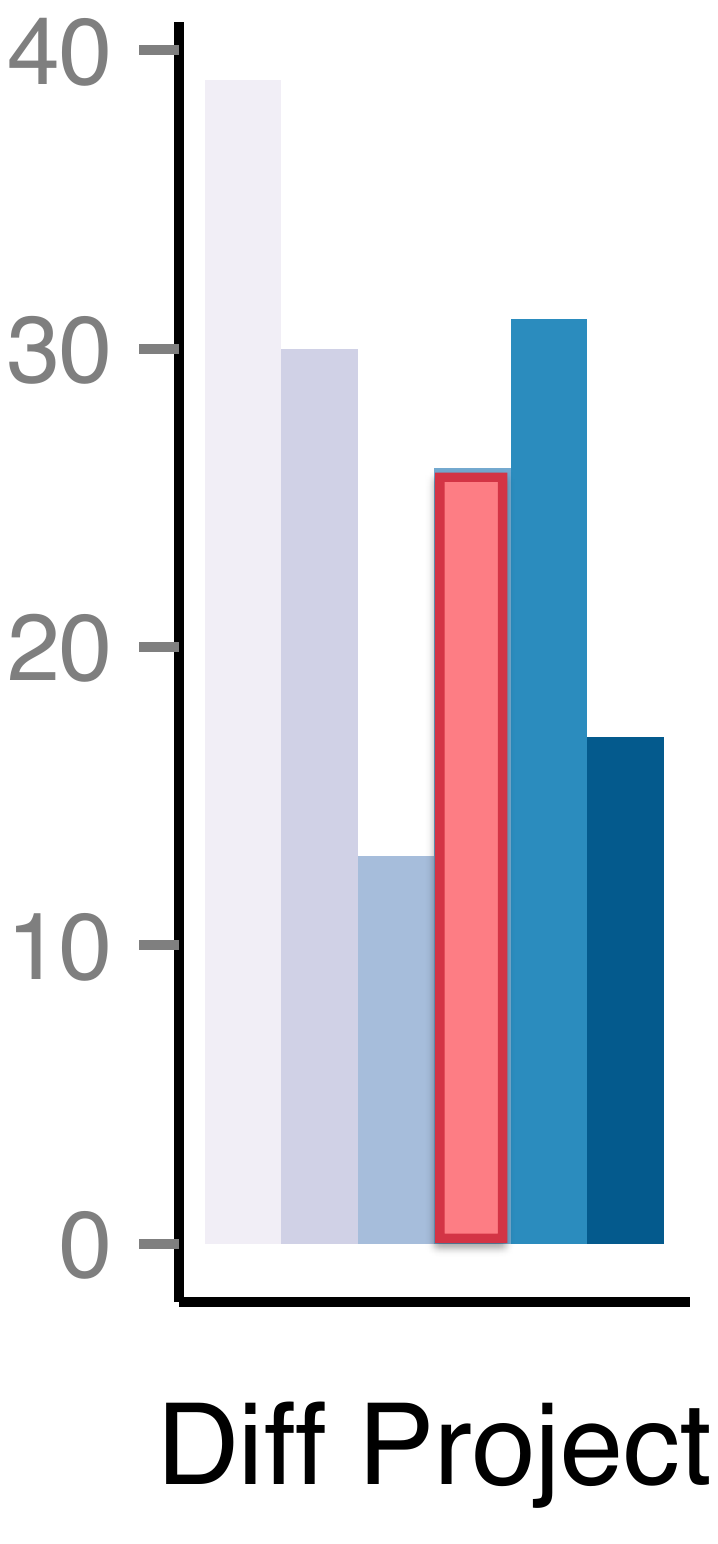}}
\subfloat{\includegraphics[scale=0.13]{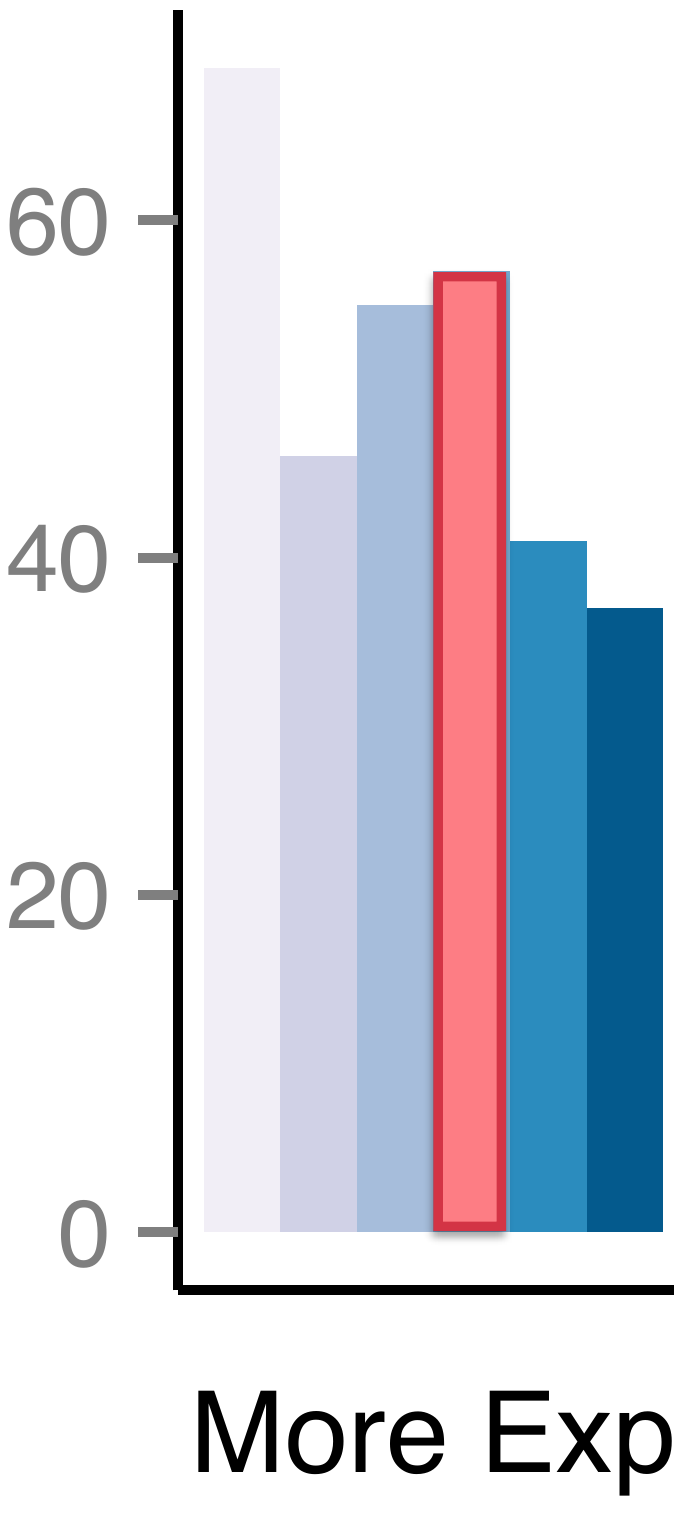}}
\subfloat{\includegraphics[scale=0.13]{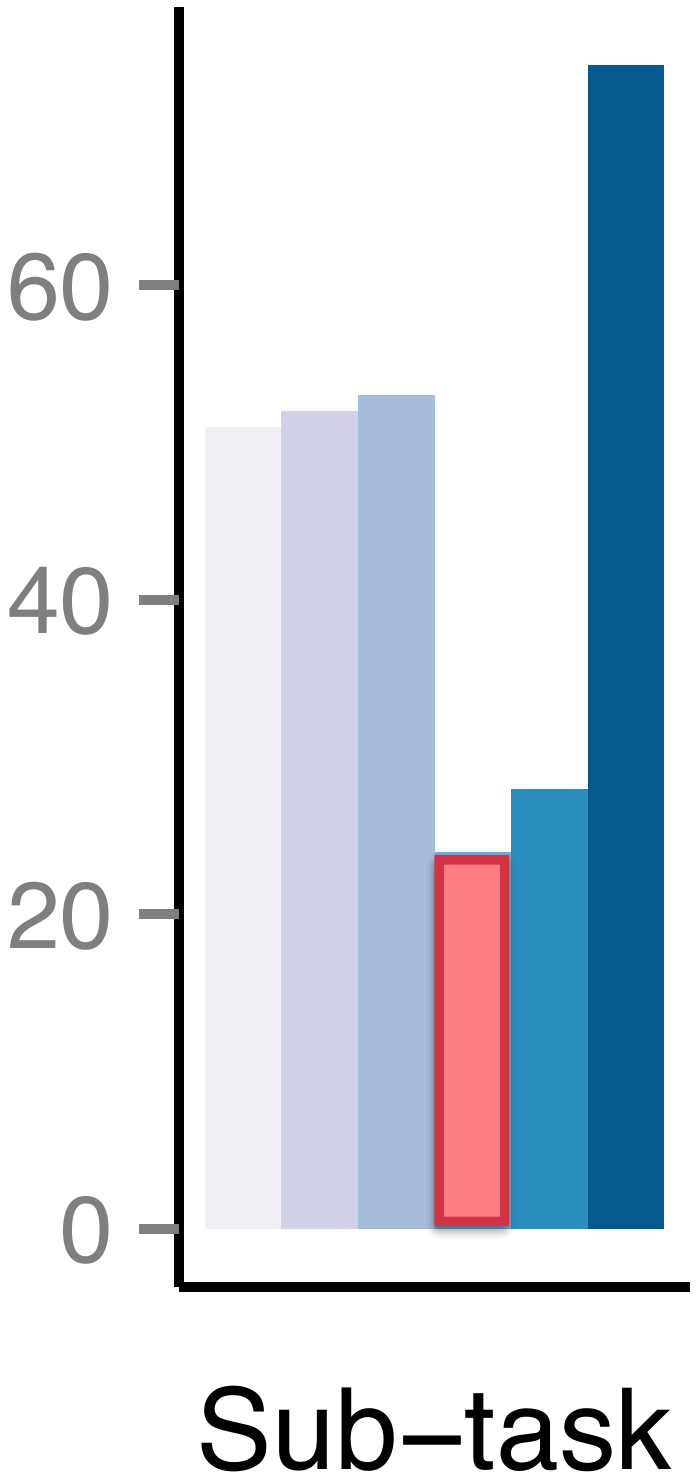}}
\subfloat{\includegraphics[scale=0.13]{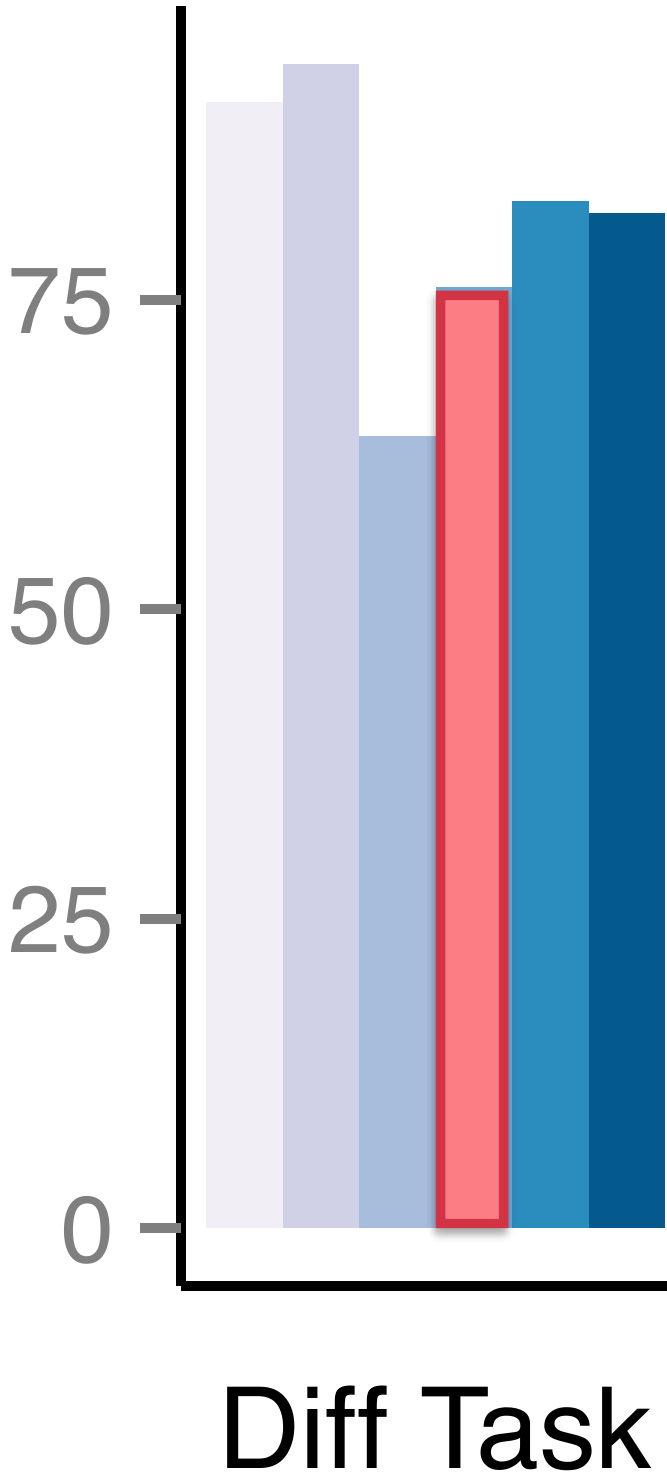}}
\subfloat{\includegraphics[scale=0.13]{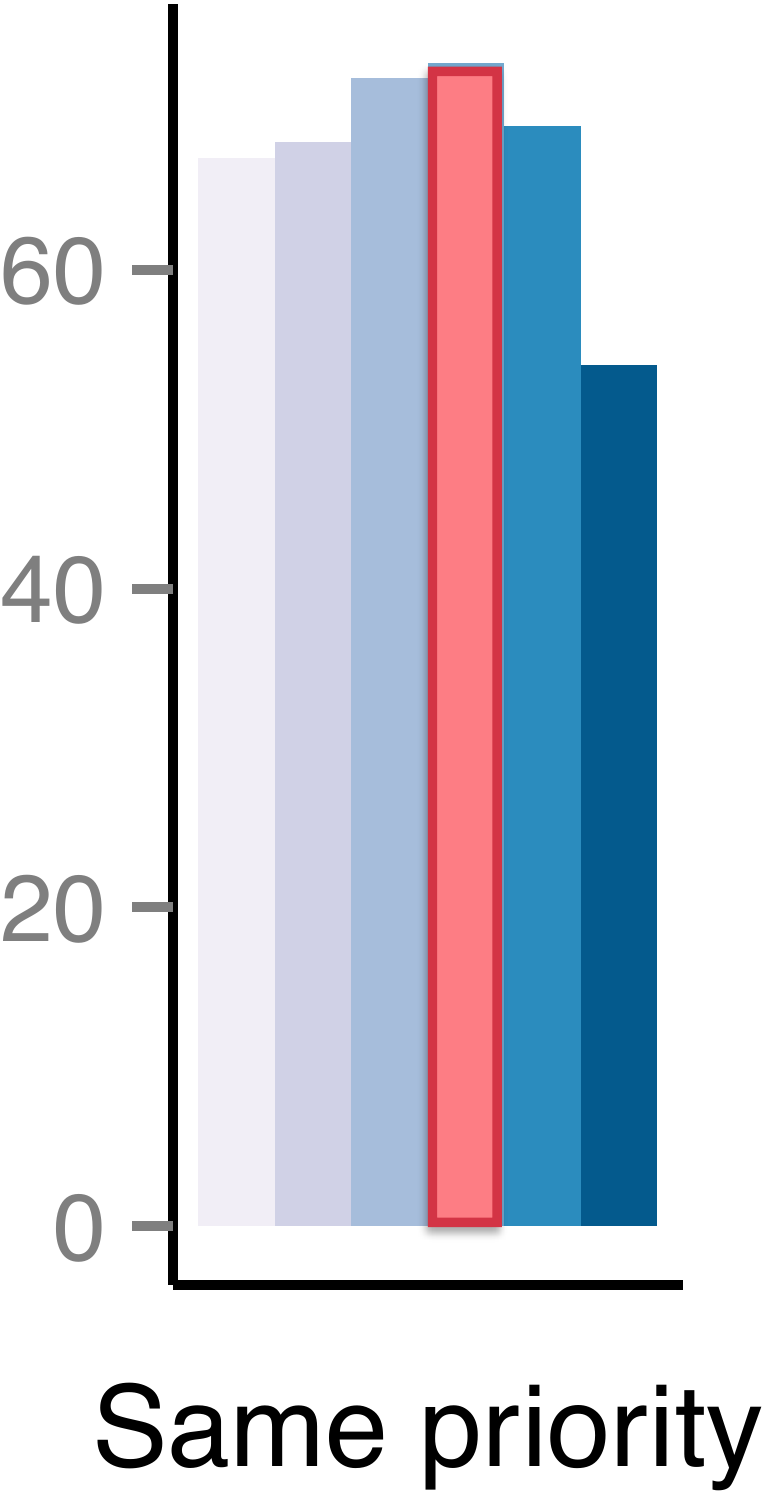}}
\subfloat{\includegraphics[scale=0.13]{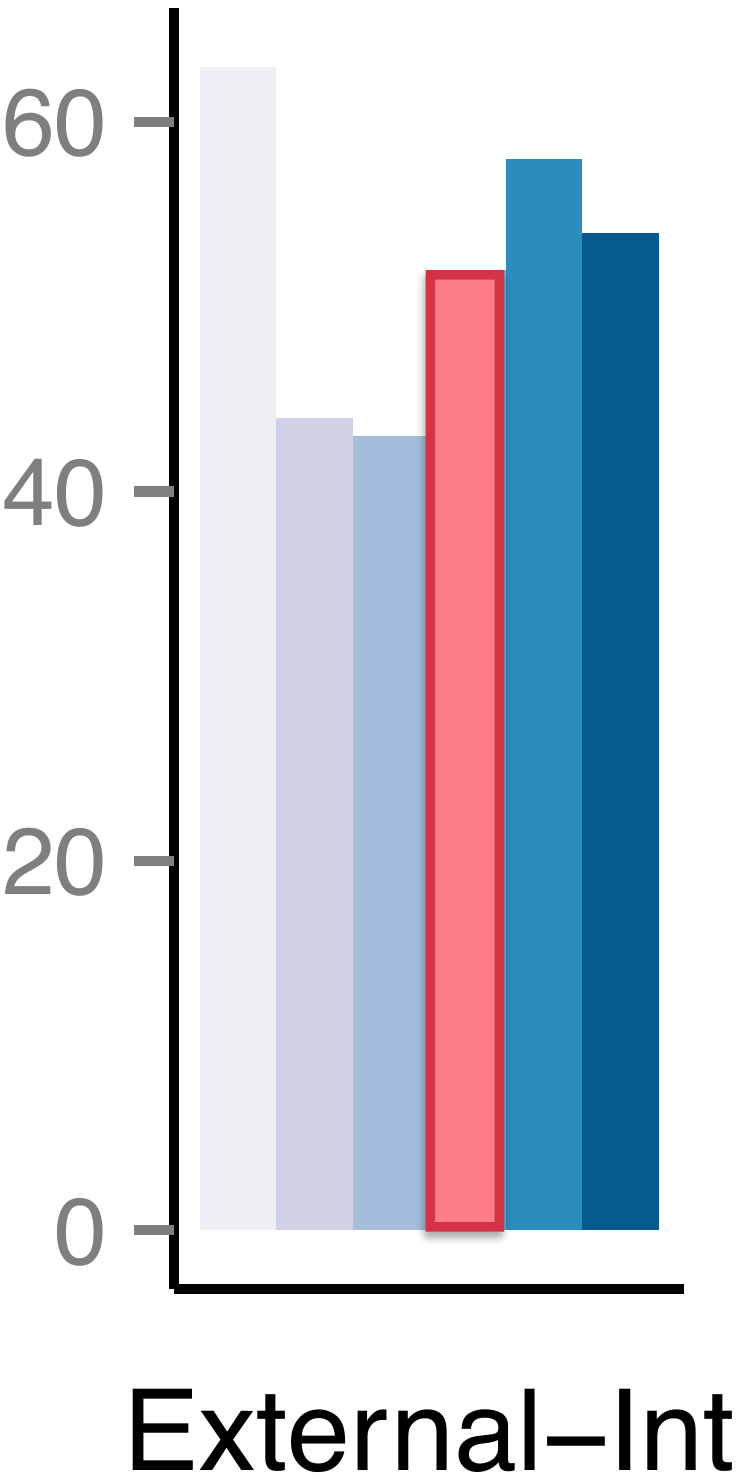}}
\subfloat{\includegraphics[scale=0.13]{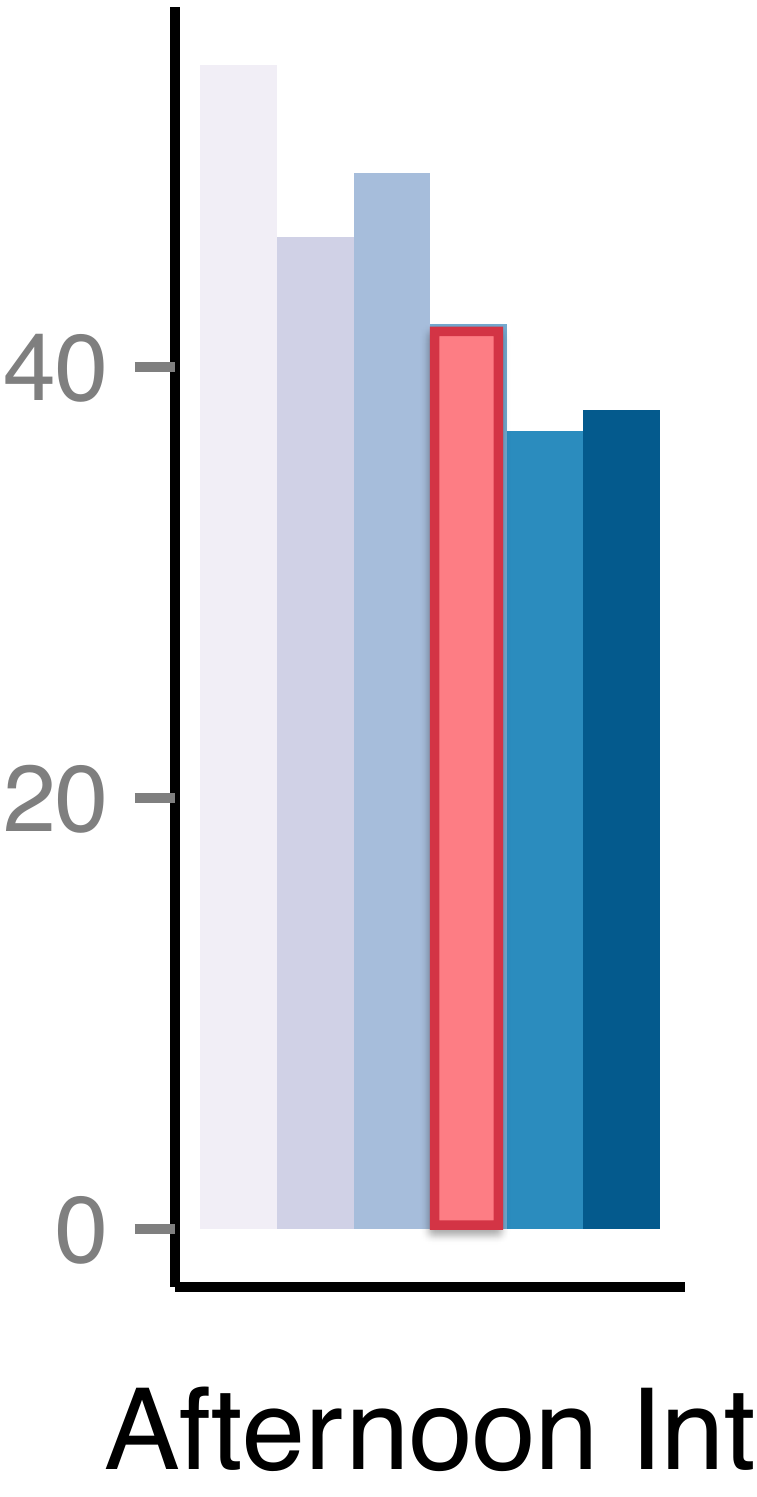}}
\subfloat{\includegraphics[scale=0.13]{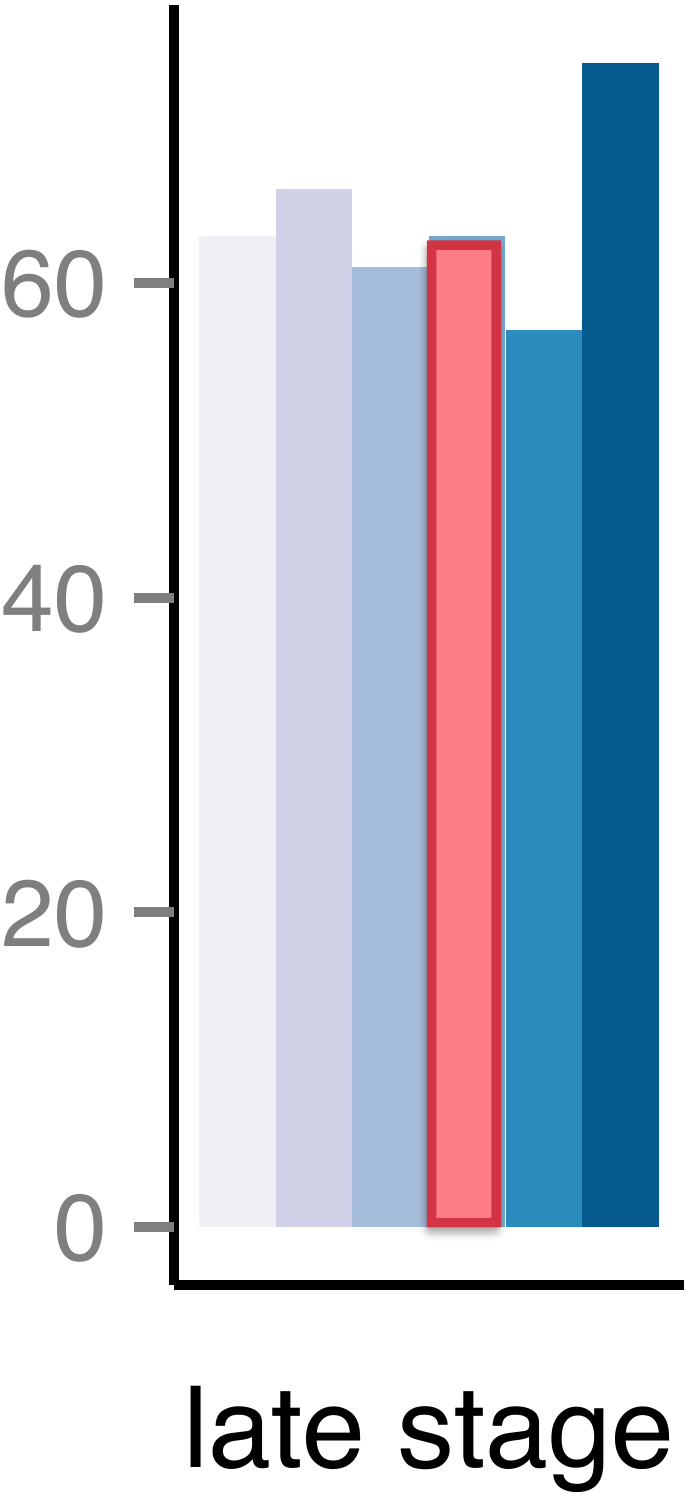}}
\subfloat{\includegraphics[scale=0.13]{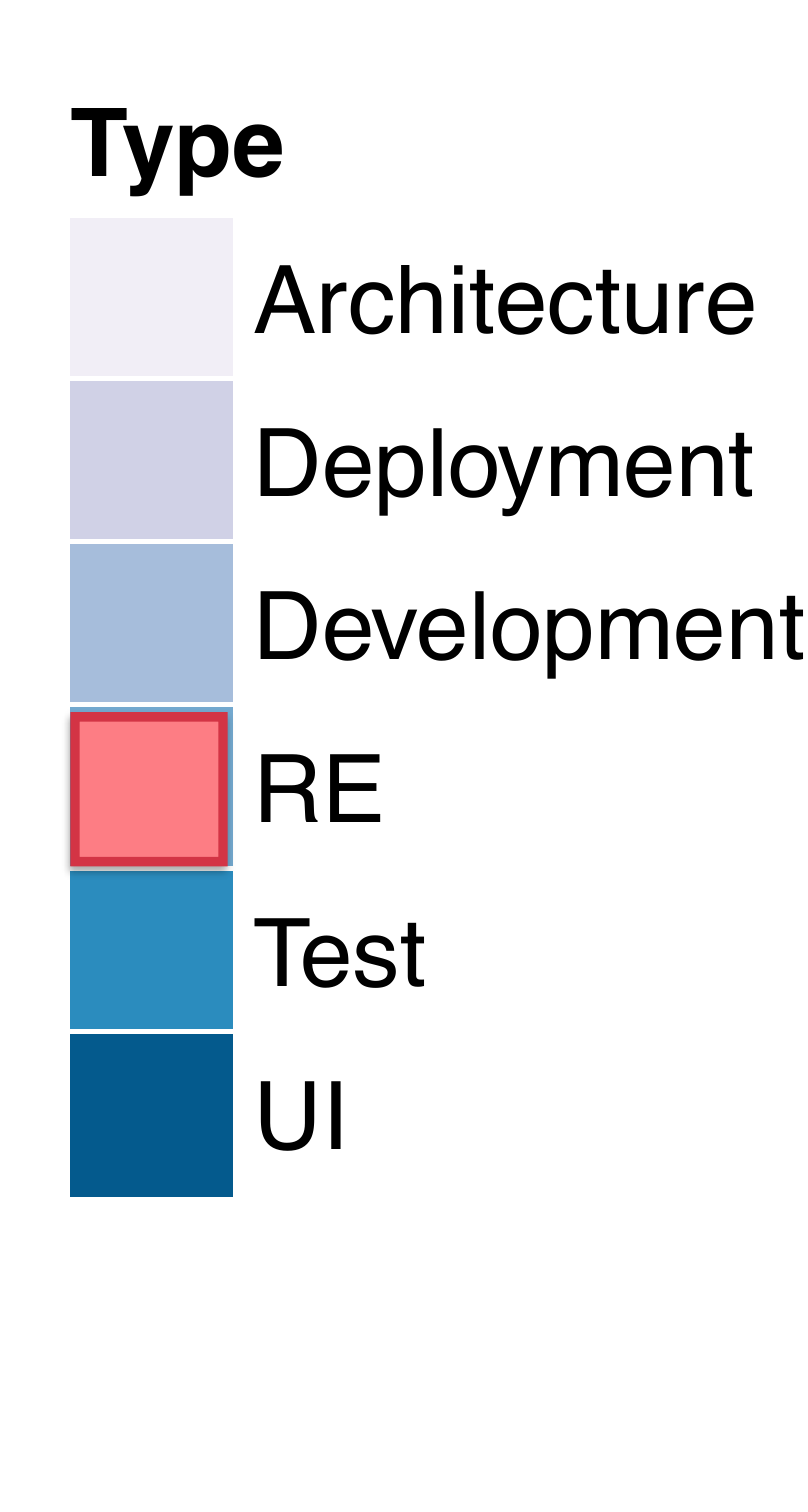}}
\vspace{-3mm}
\caption{RQ1- Comparison of frequency (percentage) of each variable between RE and other task types}
\label{fig:overall}
\vspace{-5mm}
\end{figure} 

\subsection{User Survey} 
Complementary to our retrospective analysis, we sent an online survey to 50 employees at Arcurve to understand developers' perceptions of task switchings and interruptions. We used Survey Monkey \footnote{\scriptsize http://www.surveymonkey.com} to design the
online survey and collect responses. The survey had 10 main questions (18 sub-questions) including multiple choice, Likert scale, and open-ended questions. We asked participants about their job roles, development experience in general, and the interruption factors which influence their productivity\footnote{\scriptsize https://wcm.ucalgary.ca/zshakeri/files/zshakeri/re-interruption-survey.pdf}. We received 25 (50\% response rate) responses. The average professional software development experience of participants was {12.5} (\(\pm10.5\)) years. To analyze the results of the survey, we (i) iterated through the open-ended responses using the grounded theory approach \cite{GT}, to interpret them and to identify participants' perceptions about productivity and their suggestions about the interruption tool; (ii) applied descriptive statistics and Spearman's rank correlation test methods to quantitatively investigate the survey data.
This study has been approved by the University of Calgary Conjoint Faculties Research Ethics Board (CFREB\footnote{\scriptsize http://www.ucalgary.ca/research/researchers/ethics-compliance/cfreb}).

  \vspace{-2mm}
\section{Results}
\label{sec:Results}
This section summarizes the key findings of the two-stages research, along with detailed discussion and proposed recommendations associated with each finding.

%
%
%
%
%This section summarizes key findings within three overall themes:
%In this section, the key findings of our retrospective analyses are presente
%
%This section, summarizes the key findings of the two-step research  under each RQ, 
%
%
%
%In this section, the key findings from this study are discussed and a general comparison made with the results of our survey. 
%
%the key findings of the study are first summarised and its contributions
%highlighted, then the implications for practice discussed.
\subsection{Preliminary Results}
{\bf RE activities: }According to the Kruskal-Wallis test, comparison \underline{between} RE activities showed no statistically significant differences in the disruptiveness factors we defined in Section \ref{sec:terminology} \((D_{1-3})\). Thus, in the rest of this section, we consider all RE activities as \mybox[fill=gray!30]{RE tasks}. However, the results of the related descriptive statistics (Figure \ref{fig:RETasks}) show that, to some extent, requirements elicitation is more vulnerable to task interruptions compared to other RE activities, except for the interruption lag for requirements evolution tasks. 

{\bf A productive task: } We asked each participant how they define and measure their {``productivity''} on a certain task. The participants predominantly described their productivity of a certain task as the rate of output (e.g. the number of requirements completed), as in: 
{\it  ``I define productivity of a certain task by being able to complete as much of it as possible in the shortest amount of time accurately, and with enough detail to not have to revisit it''}. 13 (54\%) participants explicitly stated that they measure their productivity as an effort spent against the expected effort. However, 4 (31\%) of these respondents stated that it is not possible to accurately estimate a task's effort and it is only possible by comparing it to previous, similar tasks.

%and in: 
%
%{\it \small ``I try to give a good estimation of the time a task will take before starting and measure my productivity (and estimation abilities) by comparing the time needed to complete the task''.} 
%

%Moreover, because of lack of sufficient data we could not apply statistical tests to statistically test the impact of each of the variables listed in Table \ref{tab:loading} on each individual RE activity. 

%Our study has shown that context determines whether interruptions are considered to be beneficial or detrimental \cite{behind}.

\begin{figure}[!t]
\centering
\vspace{-1mm}
\subfloat[Sources of External Interruptions]{\includegraphics[scale=.48]{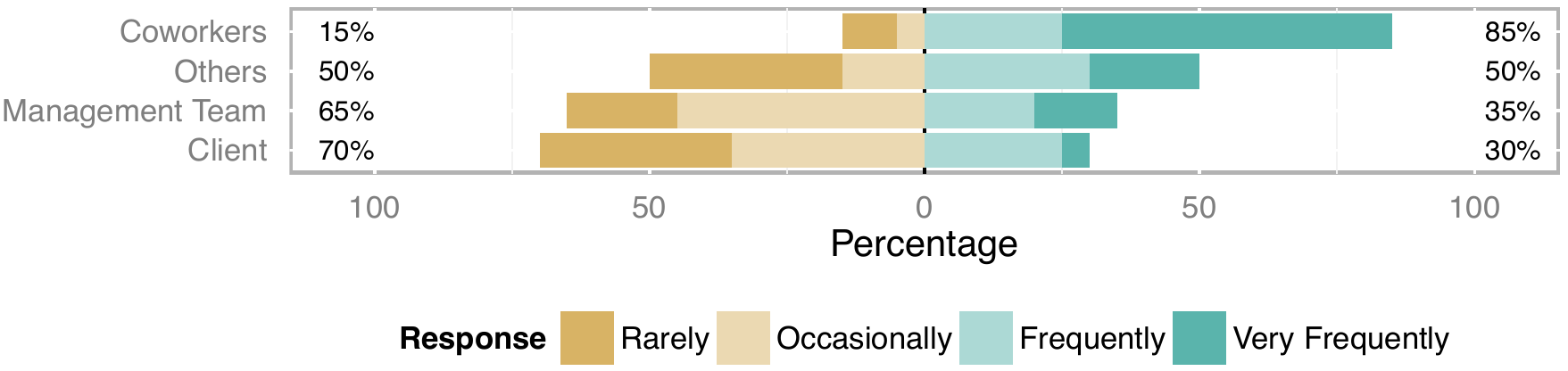}}\\ \vspace{-3mm}
\subfloat[The vulnerability of different task types to interruptions]{\includegraphics[scale=0.48]{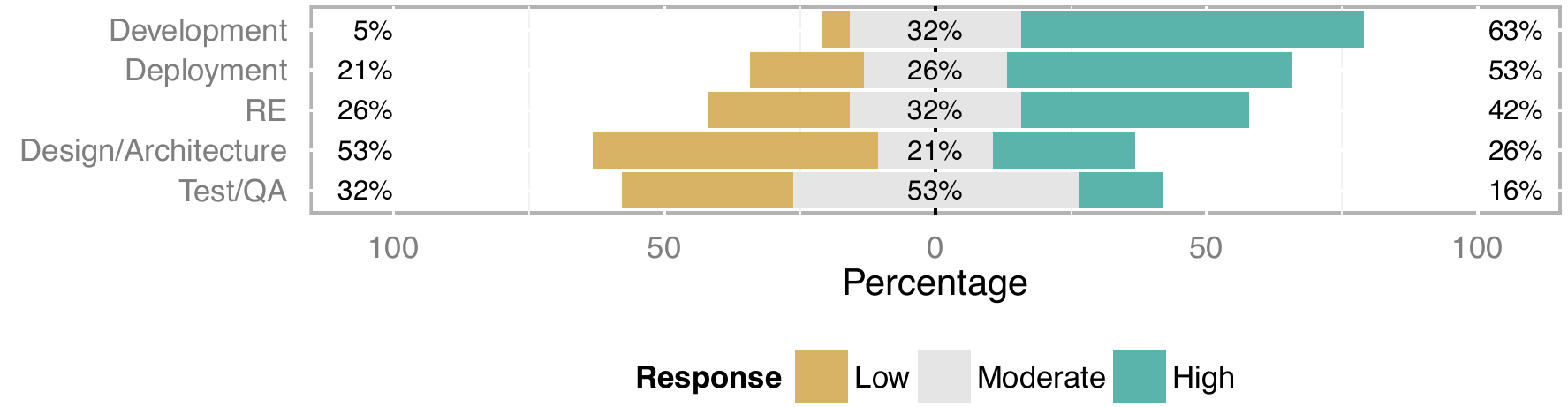}}
\vspace{-1mm}
\caption{Interruption type and source}
\vspace{-5mm}
\label{fig:Likerts}
\end{figure}

 \subsection{RQ1- RE versus Other  Software Development Tasks}

\subsubsection{Overview of Comparisons}
Figure \ref{fig:overall} illustrates the comparisons between RE and other tasks in terms of the frequency of all independent variables of this study. We observe from this figure that, compared to other types of tasks, the majority of RE and testing interruptions occur to high-level tasks. Moreover, among all RE interruptions; 24\% were sub-tasks, 26\% were caused by a different project, 57\% were carried out by employees with a higher level of experience, 76\% were caused by a different task type, 73\% were caused by a task with the same priority, 53\% were motivated by some external events, 42\% occurred in the afternoon, and 63\% occurred in later stages of their development.

%Figure \ref{fig:heatmap} (b) represents the distribution and frequency of tasks based on the duration after which they got interrupted. The median time for all of the task types we analyzed in this study is 1:00\(\pm\)00:3.

 {\bf Finding 1-1:}  We further investigated the main sources of interruptions for RE tasks. Among 53\% of external RE interruptions, 84\% are caused by coworkers, 6\% by clients, 6\% by the management team, and 4\% by other reasons (e.g. answering a phone call, or personal affairs). 

 {\bf Discussion 1-1:} Our survey data supports finding 1-1, as 85\% of participants chose {\it coworkers} as the most frequent source of their external interruptions (Figure \ref{fig:Likerts}a).

%\begin{figure}
%\begin{framed}
%\vspace*{-3.5mm}
%\centering
%\subfloat[\(v_1d_1\)]{\includegraphics[scale=0.45]{Figures/RQ1/V1D1}}
%\subfloat[\(v_3d_1\)]{\includegraphics[scale=0.45]{Figures/RQ1/V3D1}}
%\subfloat[\(v_1d_2\)]{\includegraphics[scale=0.45]{Figures/RQ1/V1D2}}
%\subfloat[\(v_2d_2\)]{\includegraphics[scale=0.45]{Figures/RQ1/V2D2}}\\[-1.8ex]
%\subfloat[\(v_3d_2\)]{\includegraphics[scale=0.45]{Figures/RQ1/V3D2}}
%\subfloat[\(v_2d_3\)]{\includegraphics[scale=0.45]{Figures/RQ1/V2D3}}
%\subfloat[\(v_3d_3\)]{\includegraphics[scale=0.45]{Figures/RQ1/V3D3}}
%\vspace*{-1mm}
%\end{framed}
%\vspace*{-3.5mm}
%\caption{{95\% confidence interval of sample means for disruptiveness of interruption characteristics in RE tasks versus other task types (Contextual characteristics)}}
%\label{fig:RQ1V1}
%\vspace{-3mm}
%\end{figure}

\begin{figure*}
\centering
\subfloat[\(v_1D_1\)]{\includegraphics[scale=0.41]{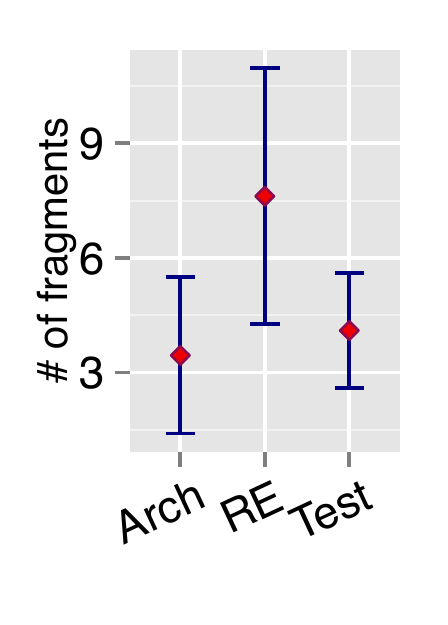}}
\subfloat[\(v_3D_1\)]{\includegraphics[scale=0.41]{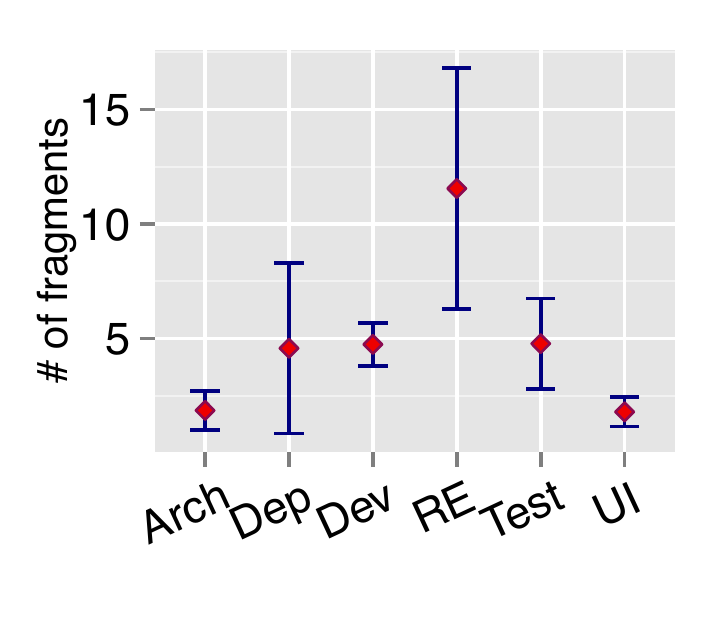}}
\subfloat[\(v_1D_2\)]{\includegraphics[scale=0.41]{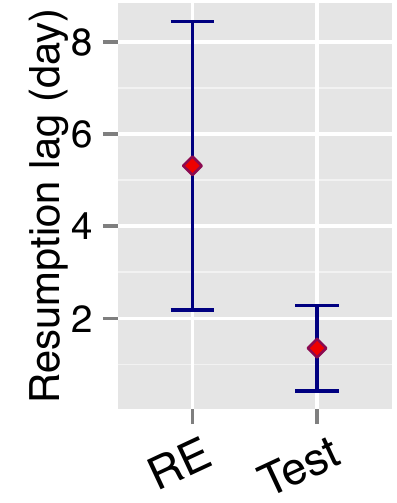}}
\subfloat[\(v_2D_2\)]{\includegraphics[scale=0.41]{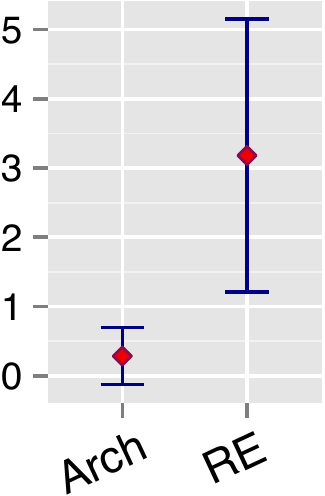}}
\subfloat[\(v_3D_2\)]{\includegraphics[scale=0.41]{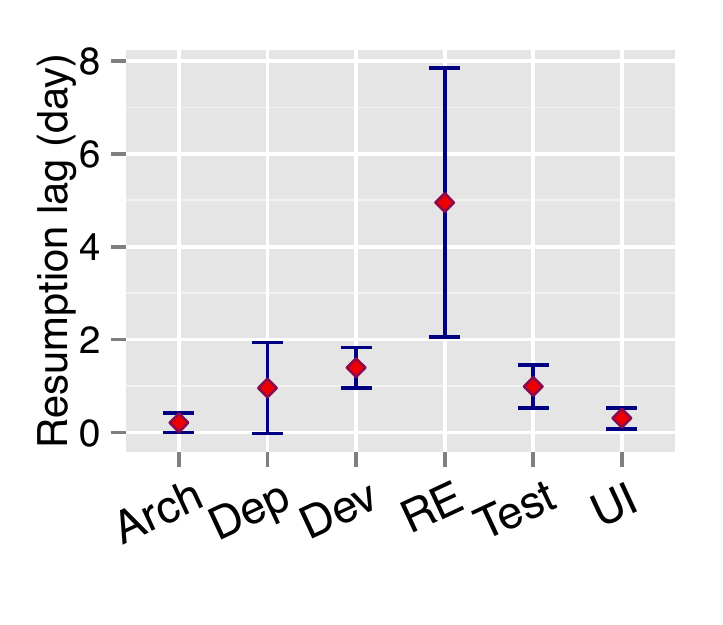}}
\subfloat[\(v_2D_3\)]{\includegraphics[scale=0.41]{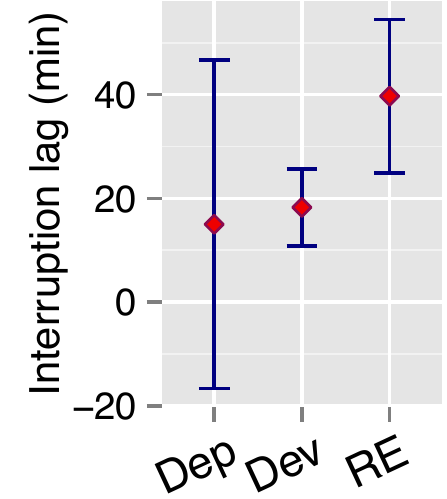}}
\subfloat[\(v_3D_3\)]{\includegraphics[scale=0.41]{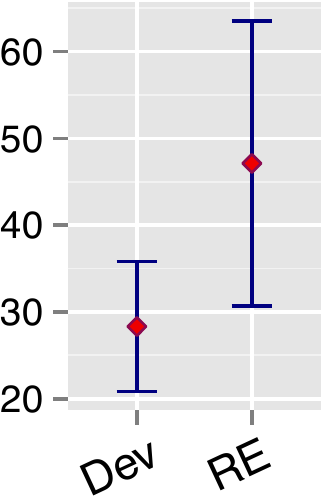}}
\subfloat[\(v_4D_1\)]{\includegraphics[scale=0.41]{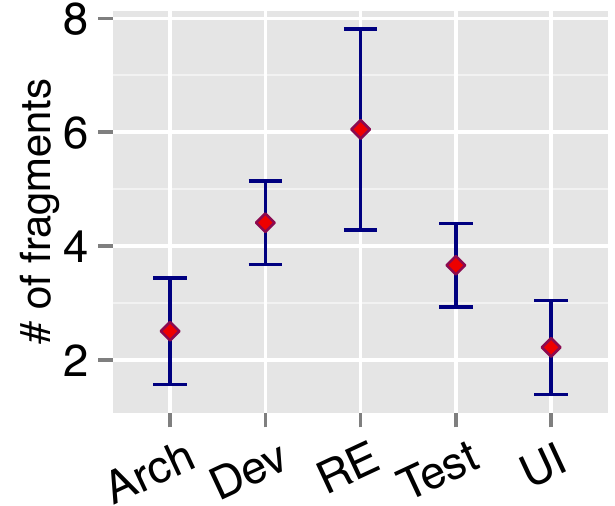}}
\subfloat[\(v_6D_1\)]{\includegraphics[scale=0.41]{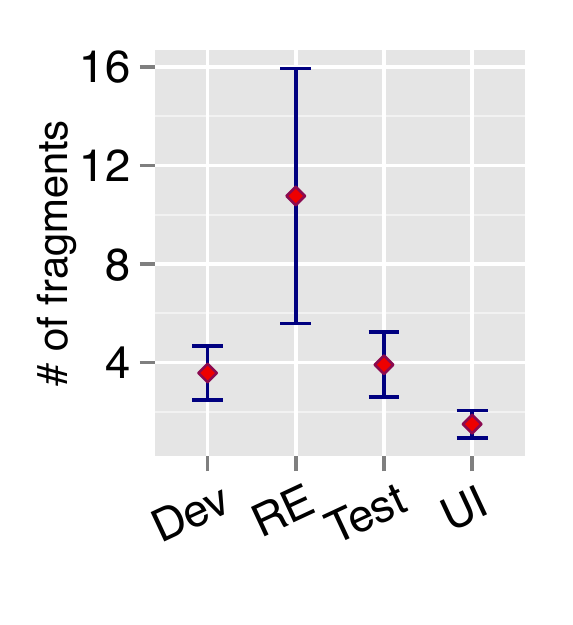}}\\[-1.7ex]
\subfloat[(\(v_4D_2\))]{\includegraphics[scale=0.41]{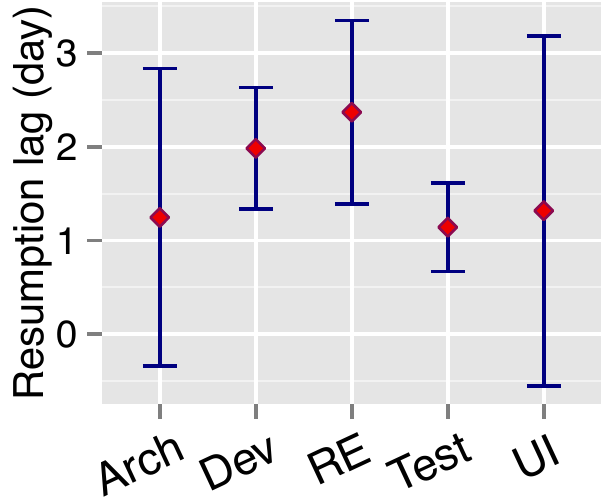}}
\subfloat[\(v_5D_2\)]{\includegraphics[scale=0.41]{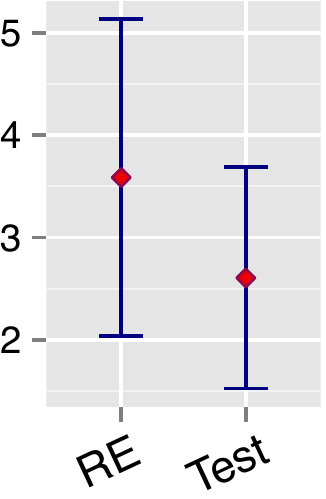}}
\subfloat[\(v_6D_2\)]{\includegraphics[scale=0.41]{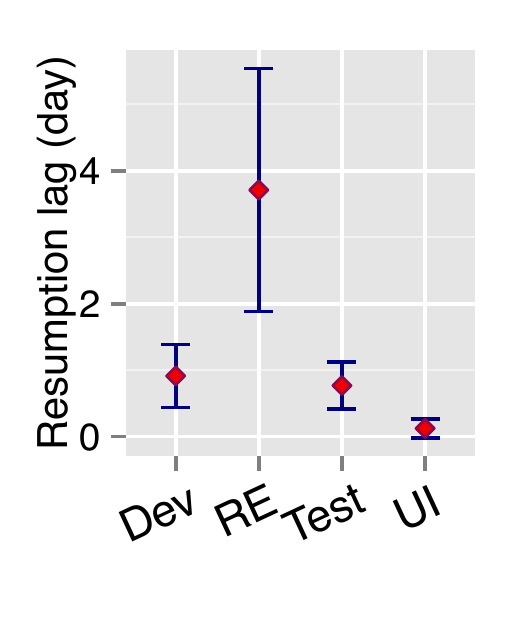}}
\subfloat[\(v_4D_3\)]{\includegraphics[scale=0.41]{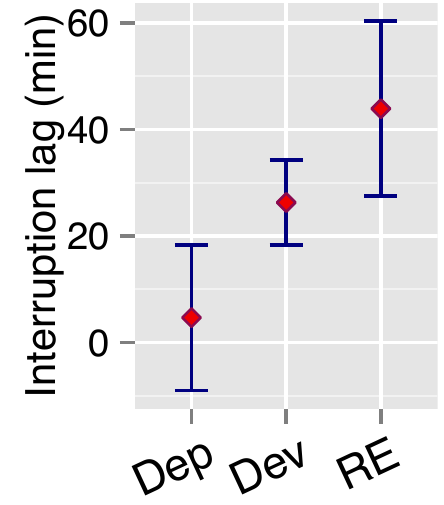}}
\subfloat[\(v_5D_3\)]{\includegraphics[scale=0.41]{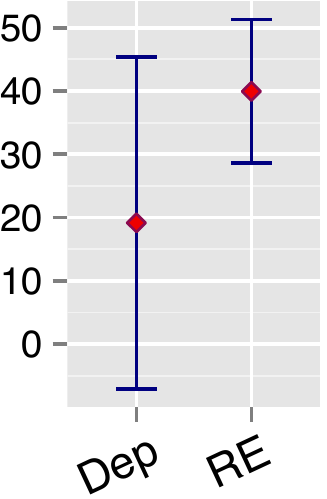}}
\subfloat[\(v_6D_3\)]{\includegraphics[scale=0.41]{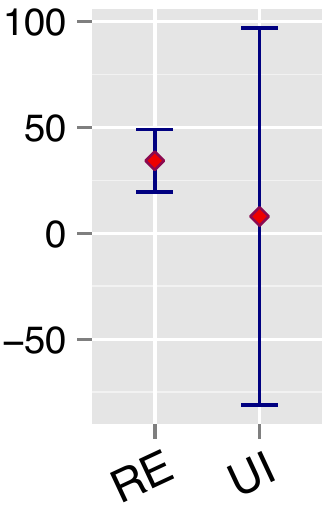}}
\subfloat[\(v_7D_1\)]{\includegraphics[scale=0.41]{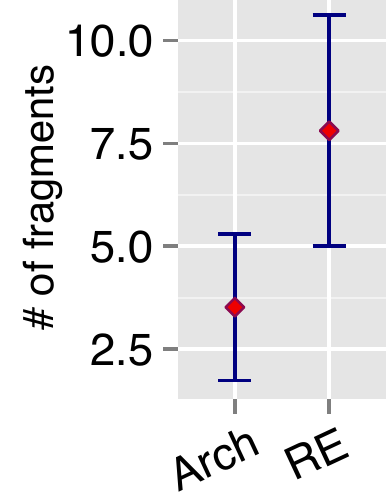}} 
\subfloat[\(v_8D_1\)]{\includegraphics[scale=0.41]{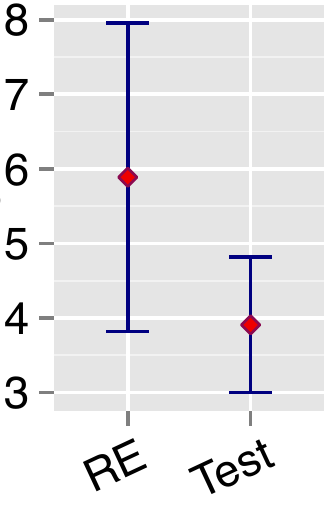}}
\subfloat[\(v_7D_2\)]{\includegraphics[scale=0.41]{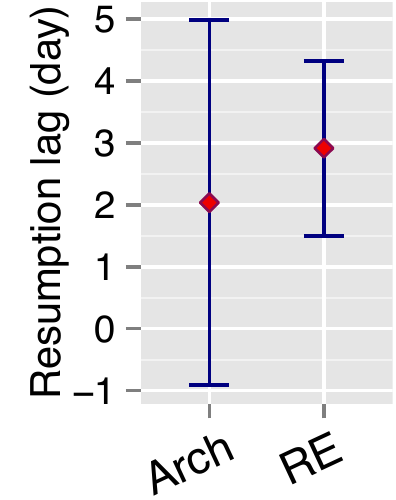}}
\subfloat[\(v_8D_2\)]{\includegraphics[scale=0.41]{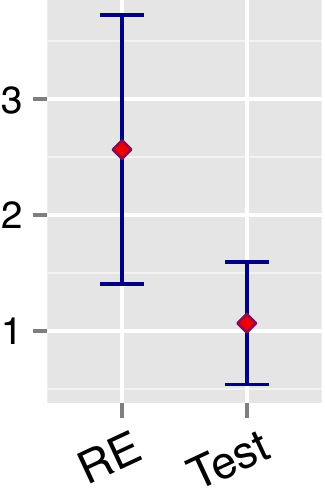}}
\subfloat[\(v_7D_3\)]{\includegraphics[scale=0.41]{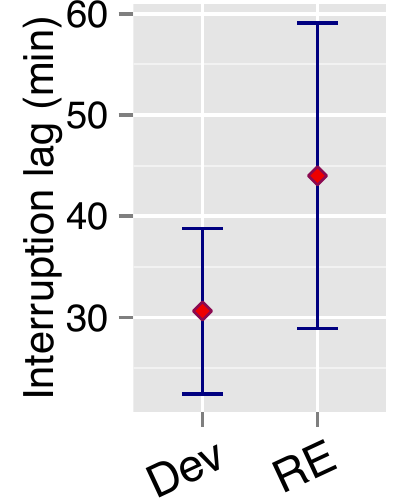}}
\caption{{95\% confidence interval of sample means for disruptiveness of interruption characteristics in RE tasks versus other task types}}
\label{fig:RQ1V1}
\vspace{-3mm}
\end{figure*}

%
%\begin{figure}
%\centering
%{\includegraphics[scale=.43]{Figures/Heatmap}}
%\vspace{-1.5em}
%\caption{RQ1 (1) Frequency of each pair of {\it (interrupting, interrupted)} tasks based on the tasks type)}
%\label{fig:heatmap}
%\end{figure} 
%
%\begin{figure}
%\centering
%{\includegraphics[scale=.5]{Figures/Duration}}
%\vspace{-1.5em}
%\caption{Distribution and frequency of interruption/task switching for different types of software development tasks}
%\label{fig:duration}
%\end{figure} 

\subsubsection{Hypothesis Testing}
to answer this RQ, we posed 144 null hypotheses following the template presented in Section \ref{sec:analysis}. Table \ref{tab:RQ1} presents the p-value for each of these tests. Out of 120 Kruskal-Wallis post-hoc tests, 41 tests were rejected, colored in gray, among which 19 (46\%) rejected tests are related to interruption characteristics (i.e. \(v_{4-6}\)), 17 (41.5\%) tests are related to the interruption context (i.e. \(v_{1-3}\)), and only 5 (12\%) of the rejected tests are related to the temporal aspects of interruptions (i.e. \(v_{7-8}\)). All of the rejected null hypothesis tests for RQ1 are illustrated in Figures \ref{fig:RQ1V1}.

 {\bf Finding 1-2:} In \underline{all} (100\%) of rejected hypotheses, RE tasks are more vulnerable to interruptions compared to other tasks. 
 
 {\bf Discussion 1-2:}  To further investigate this finding, complementary to the retrospective analysis, we asked the survey respondents to, regardless of the values of the independent variables, rank the vulnerability of main development tasks to interruptions. By vulnerability, we mean the negative impact of task switching on their productivity after resuming the primary task. As illustrated in Figure \ref{fig:Likerts}b, although 11 (43\%) participants chose RE tasks as highly vulnerable to interruptions, 16 (63\%), and 13 (53\%) found development and deployment tasks more vulnerable. This variation could be because participants answered this question according to their current job function. For example, a participant whose primary job function is {\bf RE} stated that: {\it ``I have been in situations that I have gotten requirements confused between the two tasks''}. Similarly, a {\bf tester} stated that: {\it ``I find RE is more abstract and different contexts help ensure a well thought out approach while creating a specific test bed for a test scenario takes more area specific focus or is state dependent and may have to be restarted''.} Moreover, we used different scales for these comparisons. While the retrospective analysis compares the impact of interruptions between RE tasks and other task types for specific values of our independent variables, the survey investigates these comparisons on a larger-scale, regardless of the values of the independent variables.

 {\bf Recommendation 1:} We propose the recommendation associated with this finding as a set of comparison patterns:

%\begin{figure}[!t]
%\centering
%\subfloat{\includegraphics[scale=.3]{Figures/Heatmap}}
%%\subfloat[]{\includegraphics[scale=.35]{Figures/Duration}}%#\subfloat[External Interruptions]{\includegraphics[scale=0.25]{Figures/RQ1/Triggers}}
%\vspace{-2mm}
%\caption{Interrupting type and timing}
%\vspace{-3mm}
%\label{fig:heatmap}
%\end{figure} 

\begin{tcolorbox}[colback=white, title= \textcolor{gray}{} \big< RE-\(type_k\) \big> Comparison Patterns]

{ {\textcolor{white}{\tiny ...................           }\textcolor{orange}
{\centering  Comparison Patterns: \big<{\it \((v_i D_j, RE-type_k)\)}\big>}}\\}
\begin{minipage}[t]{0.47\linewidth}
    \vspace*{0pt}
    \begin{spacing}{.75}
      \vspace{-2mm}
   {\footnotesize Table \ref{tab:RQ1} lists significantly different pairs and Figure \ref{fig:RQ1V1} presents the details of these comparisons. Moreover, Figure \ref{fig:heatmap} illustrates the percentage frequency of each pair of {\it (interrupting, interrupted)} tasks for various task types.  For instance, most of the RE tasks are interrupted by {\it development (25.7\%), PM (25.1\%)} and {\it RE (18.8\%)} tasks, respectively. Interestingly, most of the interruptions in {\it architecture (26.6\%)} and {\it UI (39.1\%)} are caused by RE tasks.}
   \end{spacing}
    \end{minipage}\hfill%
    \begin{minipage}[t]{.49\linewidth}
    \vspace*{0pt}
        \includegraphics[scale=.27]{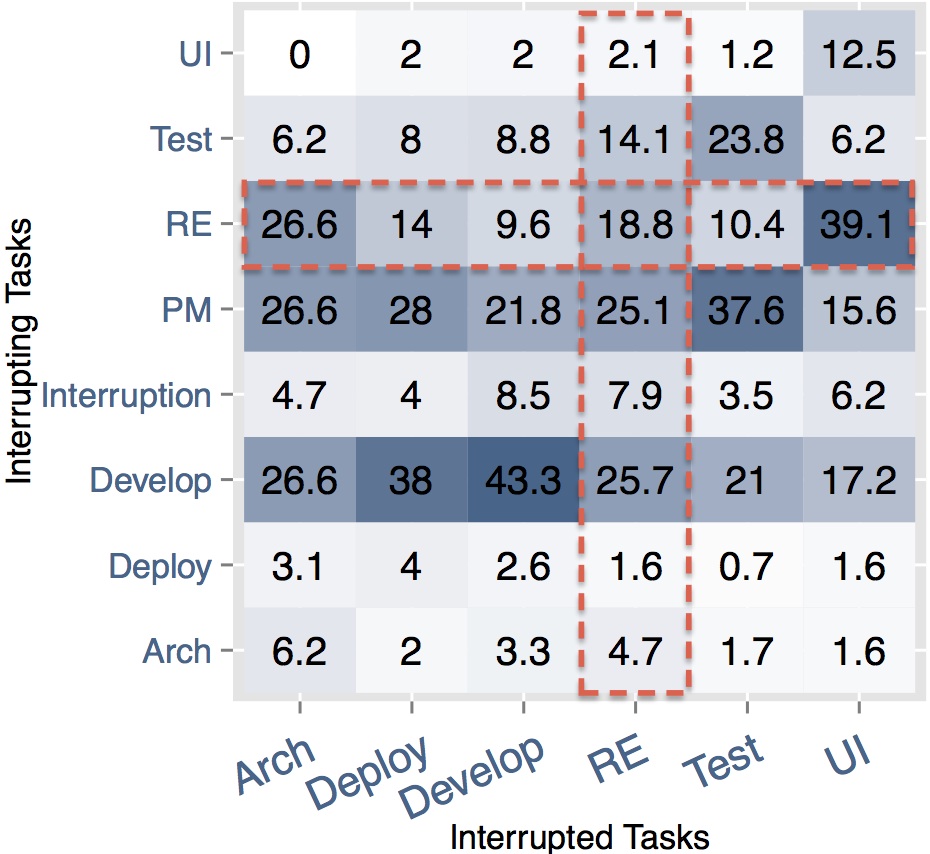}
        \vspace{-7mm}
        \captionof{figure}{\scriptsize RQ1- percentage of (interrupting, interrupted) task pairs }\label{fig:heatmap}
         
    \end{minipage}
\begin{spacing}{.9}
\vspace{-3.5mm}
\end{spacing}
\end{tcolorbox}
\vspace{-1.5mm}

%
%\begin{figure}
%\centering
%\begin{framed}
%\vspace*{-3.5mm}
%\subfloat[(\(v_4d_1\))]{\includegraphics[scale=0.5]{Figures/RQ1/V4D1}}
%\subfloat[(\(v_6d_1\))]{\includegraphics[scale=0.5]{Figures/RQ1/V6D1}}
%\subfloat[(\(v_4d_2\))]{\includegraphics[scale=0.5]{Figures/RQ1/V4D2}}
%\subfloat[\(v_5d_2\)]{\includegraphics[scale=0.5]{Figures/RQ1/V5D2}}\\[-1.6ex]
%\subfloat[(\(v_6d_2\))]{\includegraphics[scale=0.53]{Figures/RQ1/V6D2}}
%\subfloat[(\(v_4d_3\))]{\includegraphics[scale=0.53]{Figures/RQ1/V4D3}}
%\subfloat[(\(v_5d_3\))]{\includegraphics[scale=0.53]{Figures/RQ1/V5D3}}
%\subfloat[\(v_6d_3\)]{\includegraphics[scale=0.53]{Figures/RQ1/V6D3}}
%\vspace*{-1mm}
%\end{framed}
%\vspace*{-3.5mm}
%\caption{95\% confidence interval of sample means for disruptiveness of interruption characteristics in RE tasks versus other task types (Interruption characteristics)}
%\label{fig:RQ1V5}
%\vspace{-3mm}
%\end{figure}
%
%\begin{figure}
%\begin{framed}
%\vspace*{-3.5mm}
%\subfloat[(\(v_7d_1\))]{\includegraphics[scale=0.39]{Figures/RQ1/V7D1}} \hfill
%\subfloat[(\(v_8d_1\))]{\includegraphics[scale=0.39]{Figures/RQ1/V8D1}}\hfill
%\subfloat[(\(v_7d_2\))]{\includegraphics[scale=0.39]{Figures/RQ1/V7D2}}\hfill
%\subfloat[(\(v_8d_2\))]{\includegraphics[scale=0.39]{Figures/RQ1/V8D2}}\hfill
%\subfloat[(\(v_7d_3\))]{\includegraphics[scale=0.39]{Figures/RQ1/V7D3}}\hfill
%\end{framed}
%\vspace*{-3.5mm}
%\caption{95\% confidence interval of sample means for disruptiveness of interruption characteristics in RE tasks versus other task types {\bf (temporal characteristics)}}
%\label{fig:RQ1V7}
%\vspace{-3mm}
%\end{figure} 

\subsection{RQ2- RE and Interruption Characteristics}
\begin{figure}[!t]
\centering
\vspace{-2mm}
{\includegraphics[scale=.37]{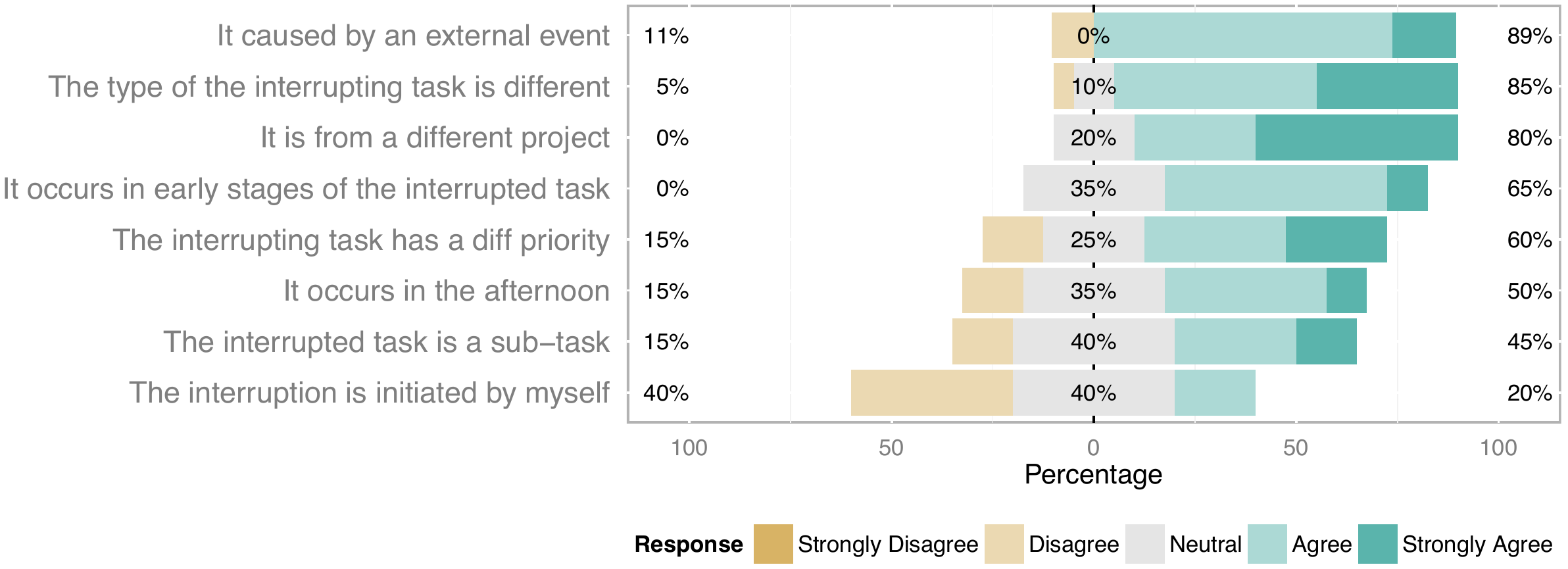}}
\vspace{-5mm}
\caption{Perceived negative impact of interruptions on RE tasks' productivity}
\label{fig:Likert}
\vspace{-5mm}
\end{figure}

To answer this RQ, we posed 24 null hypotheses following the template presented in section \ref{sec:analysis}. We also asked survey participants to indicate, using a five-point Likert-type scale, their agreement to a number of statements about the impact of each of the independent variables of our study on their productivity after RE interruptions (Figure \ref{fig:Likert}). Table \ref{tab:RQ2} presents the results of the Kruskal-Wallis tests we ran to verify all hypotheses we posed for this RQ. Moreover, we found that none of the independent variables considered in our study significantly impact the {\bf interruption lag (\(D_3\))}. In the rest of this section, we discuss the key findings and recommendations from analyzing this RQ.

%{\bf Finding 2-1:} The type and the stage of the interrupting task does not make any significant impact on the disruptiveness of RE interruptions/task switchings.\\
% \vspace{-2.5mm}
%
%{\bf Recommendations 2-1:} {\it Comparing this finding with the finding ? of RQ1, we can imply that in cases that interrupting  requirements engineer is the only option, the type of the interrupting task does not matter. However, if the same task can be assigned to other roles and its not an RE task, its better to interrupt other task types.}
% \PRLsep 
%  \vspace{-3mm}
  \begin {table*}
  \vspace{-3mm}
\centering
\caption {RQ2- The results of Null Hypothesis Testing for RQ2}
\label{tab:RQ2}
\vspace{-3mm}
\scriptsize
\begin{tabular} {|p{0.65cm}p{0.65cm}p{0.65cm}|p{0.65cm}p{0.65cm}p{0.65cm}|p{0.65cm}p{0.65cm}p{0.65cm}|p{0.65cm}p{0.65cm}p{0.65cm}|p{0.65cm}p{0.55cm}p{0.65cm}|p{0.65cm}p{0.65cm}p{0.65cm}|p{0.65cm}p{0.65cm}p{0.65cm}|p{0.65cm}p{0.65cm}p{0.65cm}|} \hline
\multicolumn{9}{|c|}{{\bf Interruption Context}}&\multicolumn{9}{c|}{{\bf Interruption Type}}&\multicolumn{6}{c|}{{\bf Interruption Time}}\\\hline

\multicolumn{3}{|c|}{{project variation}}&\multicolumn{3}{c|}{experience}&\multicolumn{3}{c|}{hierarchy}&\multicolumn{3}{c|}{{type}}&\multicolumn{3}{c|}{{self/external}}&\multicolumn{3}{c|}{{priority}}&\multicolumn{3}{c|}{{daytime}}&\multicolumn{3}{c|}{{task stage}}\\

\multicolumn{3}{|c|}{{(\(v_1 \)){\it }}}&\multicolumn{3}{c|}{{(\( v_2\))}}&\multicolumn{3}{c|}{{(\(v_3\))}}&\multicolumn{3}{c|}{{(\(v_4\))}}&\multicolumn{3}{c|}{{ (\(v_5\))}}&\multicolumn{3}{c|}{{ (\(v_6\))}}&\multicolumn{3}{c|}{{(\(v_7\))}}&\multicolumn{3}{c|}{{(\(v_8\))}}\\\hline\(D_1\)&\(D_2\)&\(D_3\)&\(D_1\)&\(D_2\)&\(D_3\)&\(D_1\)&\(D_2\)&\(D_3\)&\(D_1\)&\(D_2\)&\(D_3\)&\(D_1\)&\(D_2\)&\(D_3\)&\(D_1\)&\(D_2\)&\(D_3\)&\(D_1\)&\(D_2\)&\(D_3\)&\(D_1\)&\(D_2\)&\(D_3\)\\\hline

\cellcolor{Gray}{0.001}&\cellcolor{Gray}{0.001}&{0.5}&\cellcolor{Gray}{0.03}&{0.08}&{0.8}&\cellcolor{Gray}{0.001}&\cellcolor{Gray}{0.02}&{0.1}&{0.66}&{0.7}&{0.37}&\cellcolor{Gray}{0.0003}&\cellcolor{Gray}{1e-4}&{0.37}&\cellcolor{Gray}{0.02}&\cellcolor{Gray}{0.01}&{0.3}&\cellcolor{Gray}{0.001}&\cellcolor{Gray}{0.001}&{0.3}&{0.6}&{0.6}&{0.21}\\\hline
\end{tabular}
\vspace{-5mm}
\end{table*}

{\bf Finding 2-1 (interruption context):} According to the Kruskal-Wallis test results, in the context of RE interruptions, {project variation (\({H_0} (2, v_1D_1)\), \({H_0} (2, v_1D_2)\)), and the hierarchy level of the interrupted task (\({H_0} (2, v_3D_1)), ({H_0} (2, v_3D_2)\))} make a significant impact on the number of fragments and the resumption lag. Moreover, the experience level of a developer makes a significant impact on the number of fragments resulted from these interruptions (\({H_0} (2, v_2D_1)\)). \\
 \vspace{-2mm}

 {
{\bf Discussion 2-1:} While switching projects allows developers to use their time more efficiently, specifically in situations where they are blocked in one project, and provides them with opportunities for learning and knowledge transfer \cite{ICSE}, switching context comes at a cost. The results of our retrospective analysis also reveal that context switching increases the cognitive cost of task switching by increasing the number of task fragments and the resumption lag (Figures \ref{fig:RQ2} (a) and (g)). Our survey data also supports this finding, as 20 (80\%) participants agreed that context switching negatively impacts their productivity; {\it ``Switching gears completely affects my productivity more rather than just changing topics on the same project''}. Likewise, a recent study \cite{Zimmer} concluded that context switching generally reduces the productivity of software practitioners.
%Participants in this study believed that {\it ``being focused and Òin the flowÓ without context switches increases their productivity''.} 

Also, interruptions coinciding in low-level sub-tasks, as shown in Figures \ref{fig:RQ2} (c) and (h), are more disruptive. This was reflected in a greater number of fragments and resumption lag after switching a sub-task rather than switching a main task. The most relevant study to this finding is the one by Salvucci et al. \cite{Concurrent}, where they discussed the level of the interrupted task in terms of the complexity of the task's problem state (discussed in Section \ref{sec:terminology}) and the mental workload required for performing a task. They stated that the complexity of the problem state associated to a sub-task is higher than a main task due to having additional memory chunks. Thus, resuming lower-level tasks (e.g. requirement clarification) is more error prone and requires more effort and time. In addition, only 4 (15\%) participants disagreed about the negative impact of interrupting sub-tasks, as in: {\it ``minimal impact. Ideally, I am focused on the bigger picture''.}
%We further analyzed the impact of various levels of hierarchy on the number of fragments and the resumption lag, which yielded one of the most interesting results of this study; switching sub-tasks with lower levels of hierarchy (level three and four) was even less disruptive than switching a main task. The simplicity of the problem state associated to these tasks (e.g. ) alleviates the required effort for their resumption.  Results from an experimental study indicate that interruptions facilitate performance on simple tasks, while inhibiting performance on more complex tasks \cite{DecisionMaking}. 
\begin{figure}
\centering

\vspace{-5mm}
\subfloat[]{\includegraphics[scale=0.5]{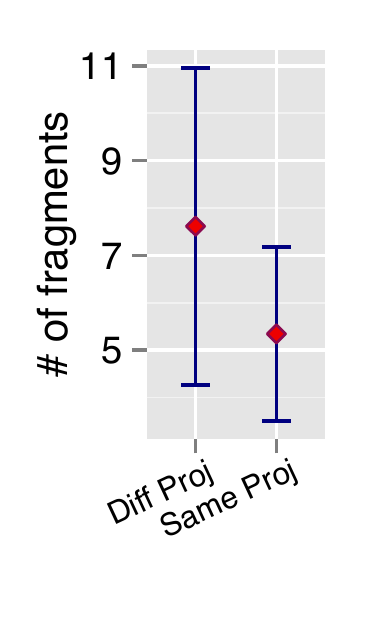}}
\subfloat[]{\includegraphics[scale=0.5]{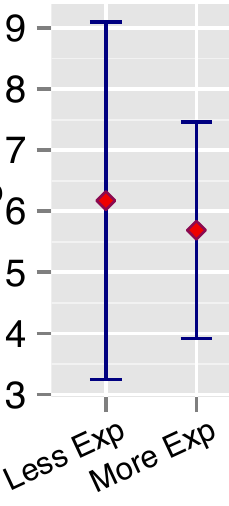}}
\subfloat[]{\includegraphics[scale=0.5]{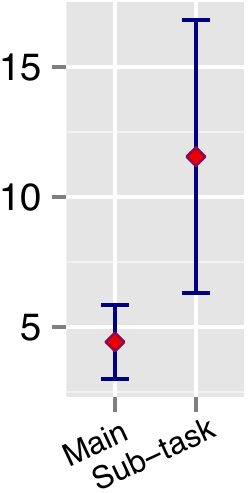}}
\subfloat[]{\includegraphics[scale=0.5]{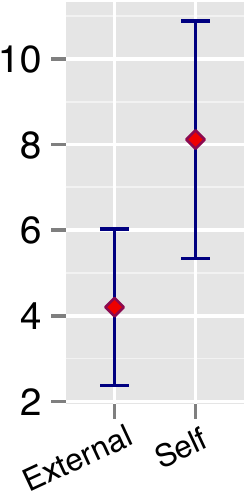}}
\subfloat[]{\includegraphics[scale=0.5]{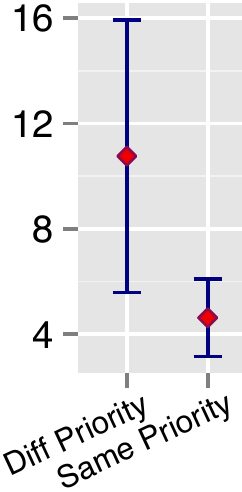}}
\subfloat[]{\includegraphics[scale=0.5]{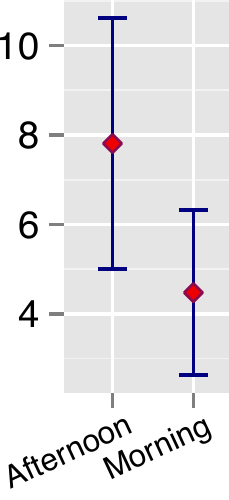}}\\[-1.95ex]
\subfloat[]{\includegraphics[scale=0.5]{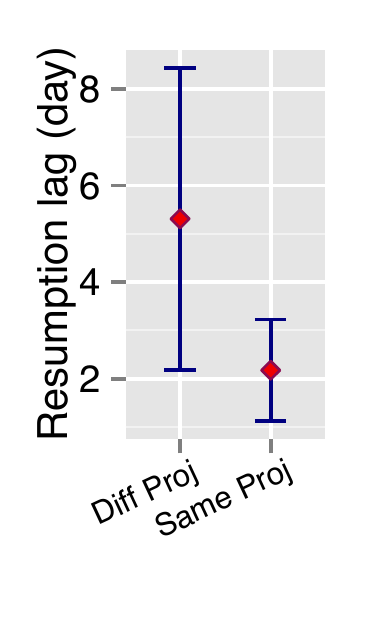}}
\subfloat[]{\includegraphics[scale=0.5]{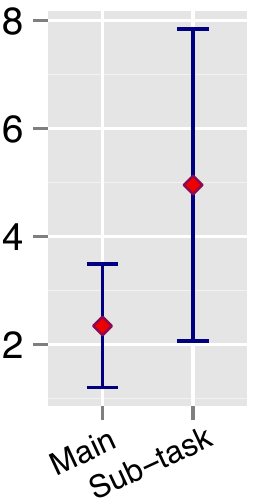}}
\subfloat[]{\includegraphics[scale=0.5]{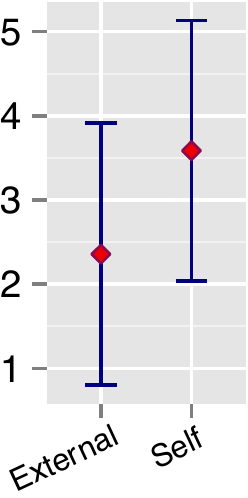}}
\subfloat[]{\includegraphics[scale=0.5]{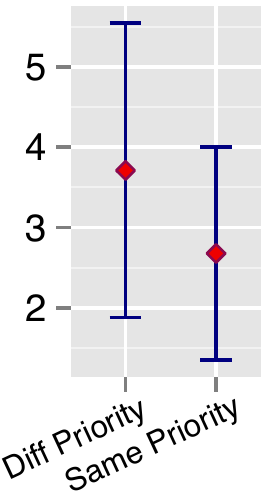}}
\subfloat[]{\includegraphics[scale=0.5]{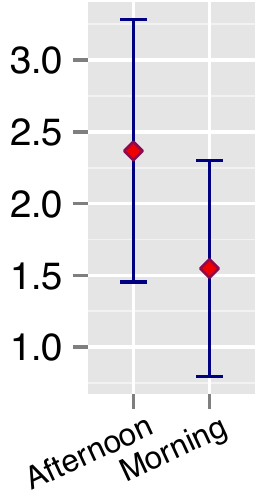}}
\vspace{-1mm}
\caption{RQ2- 95\% confidence interval of sample means for disruptiveness of interruption characteristics in RE tasks }
\label{fig:RQ2}
\vspace{-6mm}
\end{figure} 
Looking at Figure \ref{fig:RQ2} (b), we observe that interruptees with less experience resume their primary task after performing more task fragments. While our data population did not show a significant impact of experience on the length of resumption lag, we still believe that interrupting developers with less experience might have a negative impact on their productivity after resuming the task. This can be reflected in the cognitive cost of the greater number of fragments. 

{\bf Recommendation 2-1 (a):} If an RE task needs to be switched for some reasons like there is a lack of information, the interruptee is bored, or the knowledge of the interruptee is required by other teams, we recommend interrupters to possibly ask their requests from coworkers who are working on the same project as they are. Furthermore, we recommend interruptees in cases of self-interruptions to switch to a task from the same project that has a less cognitive cost of task switching. 

{\bf Recommendation 2-1 (b):} Taken together, interruptions to higher-level tasks are more desirable since they minimize cognitive costs of task switching and thus decrease the number of task fragments, resumption time and the potential for resumption errors. We propose that it might be more efficient to complete a low-level RE task before the switch.

\begin {table*}
\vspace{-5mm}
\scriptsize
\centering
\caption {RQ3- Cross-factor analysis for RE interruptions (Kruskal-Wallis Tests)}
\vspace{-3mm}
\label{tab:cross}
\begin{tabular} {p{2.3cm}|p{1cm}p{0.9cm}p{0.35cm}|p{1cm}p{0.9cm}p{0.35cm}|p{1.2cm}p{1.1cm}p{1.1cm}|p{1.2cm}p{1.2cm}p{0.5cm}|p{0.5cm}p{0.9cm}p{0.35cm}|} \cline{2-16}
&\multicolumn{3}{c}{{{same{\bf{(s)}}/ different{\bf{(d)}}}}}&\multicolumn{3}{c|}{sub/main task}&\multicolumn{3}{c|}{same {\bf (sp)}/diff ({\bf dp}) priority}&\multicolumn{3}{c|}{{morning {\bf(m)}/afternoon ({\bf a})}}&\multicolumn{3}{c|}{{early ({\bf e})/late ({\bf l}) stage}}\\

{}&\multicolumn{3}{c}{{project (\(v_1 \)){\it }}}&\multicolumn{3}{c|}{{(\( v_1\))}}&\multicolumn{3}{c|}{{(\(v_3\))}}&\multicolumn{3}{c|}{{(\(v_4\))}}&\multicolumn{3}{c|}{{ (\(v_5\))}}\\\cline{2-16}

{}&\(D_1\)&\(D_2\)&\(D_3\)&\(D_1\)&\(D_2\)&\(D_3\)&\(D_1\)&\(D_2\)&\(D_3\)&\(D_1\)&\(D_2\)&\(D_3\)&\(D_1\)&\(D_2\)&\(D_3\)\\\hline

\multicolumn{1}{|c|}{same{\bf{(s)}}/ different{\bf{(d)}}}&\multicolumn{3}{c}{}&&&&\cellcolor{Gray}0.001 {\bf (dp*)}&\cellcolor{Gray}0.002 {\bf (dp)}&\cellcolor{Gray}0.03 {\bf (dp)}&\cellcolor{Gray}0.01 {\bf(m)}&\cellcolor{Gray}0.048 {\bf(m)}&\multicolumn{1}{c|}{\ding{55}}&\multicolumn{3}{c}{}\\ 

\multicolumn{1}{|c|}{Project}&\multicolumn{3}{c}{}&&&&\multicolumn{1}{c}{\ding{55}}&\multicolumn{1}{c}{\ding{55}}&\multicolumn{1}{c|}{\ding{55}}&\cellcolor{Gray}0.005{\bf {(s)}}&\cellcolor{Gray}0.01{\bf {(s)}}&\multicolumn{1}{c|}{\ding{55}}&\multicolumn{3}{c}{}\\\cline{1-4}\cline{8-16}

\multicolumn{1}{|c|}{more{\bf{(me)}}/ less{\bf{(le)}}}&\multicolumn{1}{c}{\ding{55}}&\multicolumn{1}{c}{\ding{55}}&\multicolumn{1}{c|}{\ding{55}}&&&&\multicolumn{1}{c}{\ding{55}}&\multicolumn{1}{c}{\ding{55}}&\multicolumn{1}{c|}{\ding{55}}&\cellcolor{Gray}0.04 {\bf(m)}&\multicolumn{1}{c}{\ding{55}}&\multicolumn{1}{c|}{\ding{55}}&\multicolumn{1}{c}{\ding{55}}&\cellcolor{Gray}0.03 {\bf (e)}&\multicolumn{1}{c|}{\ding{55}}\\

\multicolumn{1}{|c|}{Experience}&\cellcolor{Gray}0.02{\bf { (me)}}&\cellcolor{Gray}0.01{\bf { (le)}}&\multicolumn{1}{c|}{\ding{55}}&&&&\cellcolor{Gray}0.02{\bf { (me)}}&\multicolumn{1}{c}{\ding{55}}&\multicolumn{1}{c|}{\ding{55}}&\cellcolor{Gray}0.003{\bf { (le)}}&\cellcolor{Gray}0.02{\bf { (le)}}&\multicolumn{1}{c|}{\ding{55}}&\multicolumn{1}{c}{\ding{55}}&\multicolumn{1}{c}{\ding{55}}&\multicolumn{1}{c|}{\ding{55}}\\\hline

\multicolumn{1}{|c|}{same{\bf{(st)}}/ different{\bf{(dt)}}}&&&&\multicolumn{1}{c}{\ding{55}}&\multicolumn{1}{c}{\ding{55}}&\multicolumn{1}{c|}{\ding{55}}&\multicolumn{3}{c}{}&\multicolumn{3}{c}{}&\multicolumn{3}{c}{}\\

\multicolumn{1}{|c|}{Task type}&&&&\cellcolor{Gray}0.001{\bf{(dt)}}&\cellcolor{Gray}0.01{\bf{(dt)}}&\multicolumn{1}{c|}{\ding{55}}&\multicolumn{3}{c}{}&\multicolumn{3}{c}{}&\multicolumn{3}{c}{}\\ \cline{1-1} \cline{5-7}
\multicolumn{15}{l}{\it *: the recorded p-value is only related to the specific values mentioned in parentheses (i.e. the \underline{controlled} variables that are used in RE Task Switching Patterns)}\\
\end{tabular}
\vspace{-5mm}
\end{table*}

{\bf Recommendation 2-1 (c):} In the context of interrupting or switching an RE task, to determine who to ask for help or more information about other tasks, we propose that it can be more efficient if interrupters ask their question to more experienced interruptees. To support this recommendation, a recent study \cite{ICSE} on task switching in software development teams, discussed the positive effects of knowledge transfers on productivity, which implies the value of experience as a source of knowledge and learning. 
%In the case of self-interruptions, we suggest interuptees with less experience to finish their current task before switching to another task. In cases that interruption is inevitable, the application of resumption strategies such as stick notes or digital reminder might help with reducing the cognitive cost of task fragments. 

 \PRLsep 
  \vspace{-3.65mm}
  
     {\bf Finding 2-2 (interruption characteristics):} According to the Kruskal-Wallis test results, in the context of RE task switchings and interruptions, the {source of the interruption (\({H_0} (2, v_5D_1)\), \({H_0} (2, v_5D_2)\)), and the interrupted task priority (\({H_0} (2, v_6D_1), {H_0} (2, v_6D_2)\))} make a significant impact on the number of fragments and the resumption lag. \\
 \vspace{-3.5mm}

{\bf Discussion 2-2:} As shown in Figure \ref{fig:RQ2} (d), and (i), self-interruptions in the context of RE interruptions cause more task fragments and longer resumption lags. This finding is in line with the results of a controlled experiment that Katidioti et al. \cite{selfinterruption} conducted to contrast external interruptions with self-interruptions. Their study shows that self-interruptions introduce the extra cost of the decision. If these extra costs do not lead to a substantive reduction in the other costs of interruption, self-interruptions are more harmful than external interruptions. This is because they lead to greater resumption lags and have a negative impact on the duration of the interrupted task. However, our survey participants were not very positive about the effects of external interruptions on their productivity: only 9\% agreed that self-interruptions negatively impacts their productivity, while this number for external interruptions is 89\% (Figure \ref{fig:Likert}). Moreover, 20 (80\%) participants indicated that less than 30\% of their interruptions are self-interruptions. To further investigate this difference, we hypothesized that there might be a correlation between the number of projects that participants are involved in and their response to this question. A Spearman's rank correlation test shows no correlation between the participants' agreement to the impact of self/external interruptions and  the ``\# of projects'' they are involved in. Also, there is no correlation between the size of the primary project (in people) of participants and their agreement to these statements. However, there was a strong correlation between the participants' agreement to the negative impact of \(\big<\)context switching, task type\(\big>\), and self-interruption (context switching: rho = 0.61, {\it p} = 0.01; task type: rho = 0.5, {\it p} = 0.03). In addition, the Spearman's rank correlation test shows no correlation between the participants' agreement to the negative impact of external interruptions, the negative impact of context switching and type of the interrupting task. 

Interestingly, according to the Kruskal-Wallis test results, the type of the interrupting task (\(v_4\)), does not make any significant impact on the disruptiveness of interruptions in RE tasks ({\it p-value\(>\)0.05}). {However, 21(85\%) survey participants agreed that their productivity will be negatively impacted if the interrupting task is a task with a different type (e.g. development, test); {\it ``Tool/environment switching seems to burn more time than I usually account for''.}} We aim to further investigate this variable in our planned replication study. 

%Mark et al. \cite{behind} studied the impact of self and external interruptions on task switching and the interruption timing. They found that managers were more likely to experience external interruptions (59.2\%) than self-interruptions (40.8\%), whereas this was about equal for other work roles such as analysts and developers. They also found that about half (53.3\%) of the externally interrupted tasks are resumed on the same day.

{\bf Recommendation 2-2:} We propose that if an RE task needs to be switched, it can be more efficient if this switch is triggered by an external source such as coworkers or the management team, instead of leaving the decision to the interruptee. 
 \PRLsep 
  \vspace{-4mm}
  
{\bf Finding 2-3:} according to the Kruskal-Wallis test results, in the context of RE interruptions, daytime (\({H_0} (2, v_7D_1)\), \({H_0} (2, v_7D_2)\)) makes a significant impact on the number of fragments and the resumption lag. 

 \begin{figure}
\centering
\vspace{-2mm}
{\includegraphics[scale=.3]{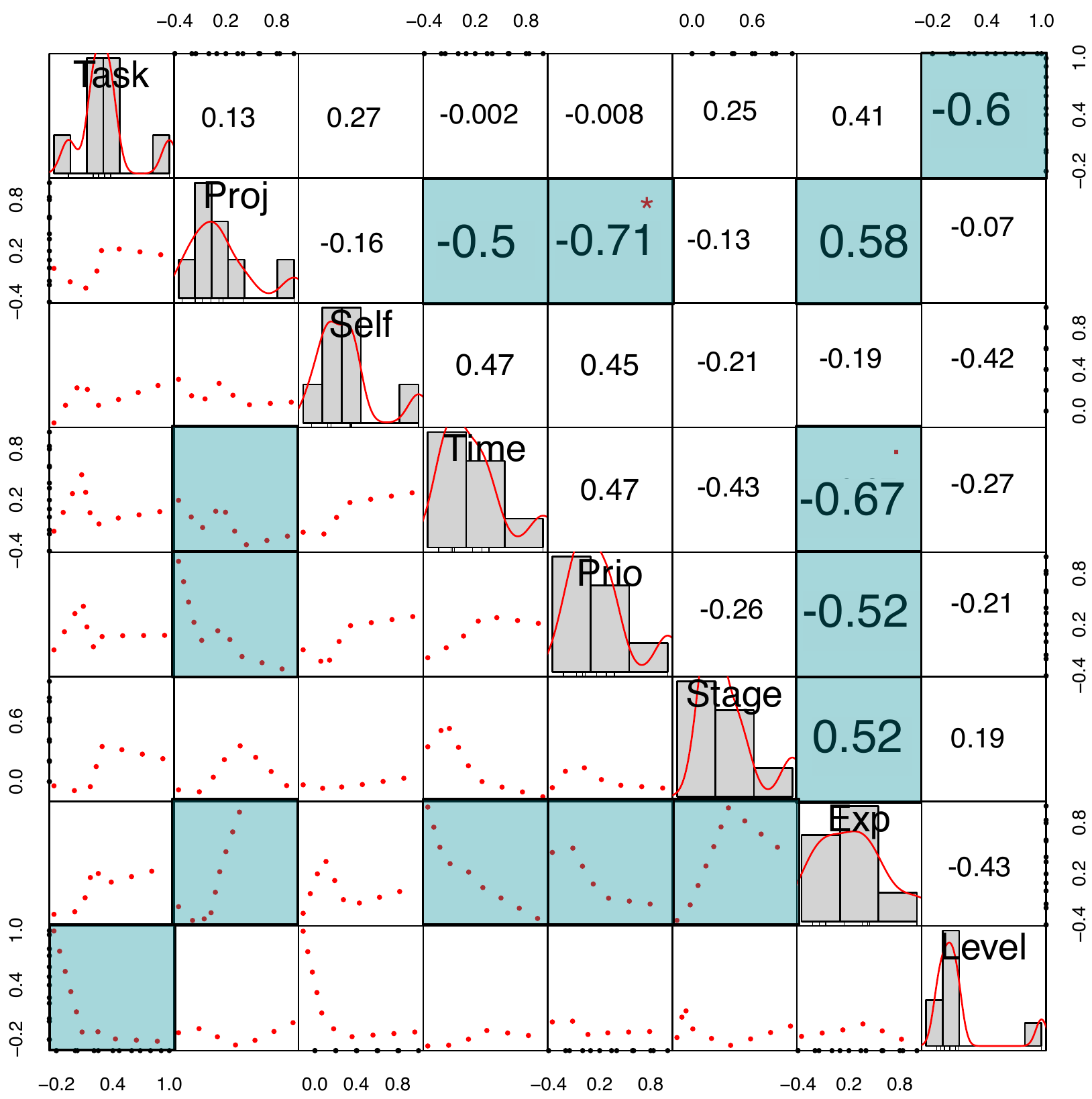}}
\vspace{-1em}
\caption{Cross-correlation between \(v_{1-8}\) for interrupted RE tasks}
\vspace{-5mm}
\label{fig:cross}
  \vspace{-1mm}
\end{figure} 

{\bf Discussion 2-3:} As shown in Figure \ref{fig:RQ2} (f), and (k), in the context of RE interruptions, afternoon interruptions cause more task fragments and longer resumption lags. Of all recorded RE interruptions, 57\% were interrupted in the morning among which 63\% were resumed on the same day, while this number for afternoon interruptions is 39\%. Similarly, our results show that about half (56\%) of the afternoon interruptions are self-initiated. This might be due to the additional pressure for completing more tasks before the end of the day. Further, 12 (50\%) survey participants agreed and only 4 (15\%) disagreed about the negative impact of afternoon interruptions, as in: {\it ``if I get interrupted later in the day I tend to use the existing interruption to get a coffee as opposed to waiting until I finish a current task''}. Also, a Spearman's rank correlation test shows a strong correlation between participants' agreement to the negative impact of afternoon interruptions and the size of the project they are involved in (rho= 0.5, {\it p} = 0.04).

Although none of the survey participants disagreed about the negative impact of early-stage interruptions on their productivity (i.e. 65\% (strongly) agreed, 35\% neutral), our retrospective analysis does not show any significant impact of this variable (\(v_8\)) on RE interruptions (i.e. all {\it p-values\(>\)0.05}), as in: {\small \it ``the early thought process in new development are key. If I swap in the middle of this, it will take longer to recreate the thought process since it is less developed mentally''.} On the other hand, several studies (e.g. \cite{Instant, Temporal}) measured the cognitive cost of interruptions by looking at the point at which a task is interrupted. They found that interrupting a task during middle or end-task interruption points will result in longer interruption and resumption time. This implies the need for a further investigation on the impact of interruption point on RE interruptions.  

%A recent observational study \cite{behind} stated that practitioners typically postpone tasks that demands more concentration towards the end of the day due to the less number of interruptions they receive from their coworkers. 
%

{\bf Recommendation 2-3:} Given the negative impact of self-interruptions on interrupting RE tasks (Finding 2-2) and the cost of afternoon interruptions, we recommend practitioners to minimize voluntary task switchings in later times of a day.

\subsection{RQ3- Cross-factor Analysis} 
To answer this RQ, we conducted a cross-factor analysis of all of the variables listed in Table \ref{tab:loading} in the context of RE task interruptions. As the first step, we performed a cross-factor correlation analysis among all of these variables (Figure \ref{fig:cross}) and excluded all the unsubstantial cross-factor combinations (i.e. \(r< 0.5\) \cite{correlation}). Out of all possible 28 combinations, we selected 7 combinations for further analysis, which are highlighted in Figure \ref{fig:cross}. As the next step, we tested the significant impact of each of these combinations on RE tasks (Table \ref{tab:cross}). To properly interpret the cross-factor impact of these combinations, we used a combination of Violin, Scatter, and Box plots for each of these combinations (e.g. Fig \ref{fig:Violin}). 

 {\bf Finding 3 and Recommendation 3:} We formulate the main findings of this RQ in forms of some {\bf RE task switching patterns}. In the rest of this section, we briefly list (Table \ref{tab:cross} and Figure \ref{fig:cross}) and discuss these patterns. 

% This introduces some space you can remove or change it.
 % An example use
{\bf Discussion 3: }
10 (47\%) participants agreed to the negative impact of both interruptions with different type, and interrupting sub-tasks [P4]. 11 (79\%) and 10 (71\%) of more experienced participants agreed with the negative impact of early-stage interruptions [P6] and context switching [P3], respectively. A participant with 12 years of experience stated: 
{\it ``When formulating initial ideas and solutions, I find when I am grouping multiple thoughts and they haven't formed a concrete assertion, if interrupted, I have to start over. Once I have a generalized thought, I tend to resume quickly.''} 

\vspace{-1mm}
\begin{tcolorbox}[colback=white, title= RE Task Switching Patterns]
\vspace{-2mm}
{ {\textcolor{white}{\tiny ..................................           }\textcolor{orange}{\centering \big<{\it (\(V_{ctl}, V_{ind}, \text{\it Val} _{dis}\)), \(D_{1-i}\)}\big>} {\small\(i \in \{1, 2, 3\}\)}}
{\small
Where \(V_{ctl}\) and \(V_{ind},\) denote the controlled and independent variables  and \(\text{\it Val}_{dis}\) represents the value of the independent variable which makes an RE task switching more disruptive. Moreover, \(D_i\) represents the disruptiveness measures.}}\\
\begin{spacing}{.9}
\vspace{-2mm}
\textcolor{orange}{\small{\it{{Example: \big<{\it (same project, daytime, afternoon), \(D_{1-2}\)}\big>  states that  when switching RE tasks on the same project, ``daytime'' makes a significant impact on the disruptiveness of these interruptions, and \underline {afternoon} interruptions have a greater number of fragments and resumption lag comparing to task switchings that occur during the morning. In the case of context switching, {\it daytime} does not make a significant impact on the disruptiveness of RE task switchings [P1]. \\}}}}
\end{spacing}
\vspace{-2mm}
{\underline{\small \bf Patterns} 

\noindent

{\bf \small [P1] } {\small{\it{{\big<{\it (same project, daytime, afternoon), \(D_{1-2}\)}\big>}}}}

{\bf \small [P2]  }  {\small \textcolor{white}{}\big<{\it (morning, project variation, different project), \(D_{1-2}\)}\big>} 

{\bf \small [P3] }  {\small \textcolor{white}{}\big<{\it (more exp, project variation, different project), \(D_{1}\)}\big>}

{\bf \small [P4] }  {\small \textcolor{white}{}\big<{\it (different type, task level, sub-task), \(D_{1-2}\)}\big>}}

{\bf \small [P5] }  {\small \textcolor{white}{}\big<{\it (same priority, project variation, diff project), \(D_{1-3}\)}\big>}

{\bf \small [P6] }   {\small \textcolor{white}{}\big<{\it (early stage, experience, more experience), \(D_{2}\)}\big>}

{\bf \small [P7] }   {\small \textcolor{white}{}\big<{\it (less experience, project variation, diff project), \(D_{2}\)}\big>}

{\bf \small [P8] }   {\small \textcolor{white}{}\big<{\it (morning, experience, more experience), \(D_{1}\)}\big>}

{\bf \small [P9] }   {\small \textcolor{white}{}\big<{\it (more experience, priority, different priority), \(D_{1}\)}\big>}

{\bf \small [P10]}   {\small \textcolor{white}{}\big<{\it (less experience, daytime, afternoon), \(D_{1-2}\)}\big>}

\vspace{1mm}
\begin{minipage}[t]{0.001cm}
    \vspace*{0pt}
    \end{minipage}\hfill%
    \begin{minipage}[t]{8cm}
\centering
        \includegraphics[scale=.62]{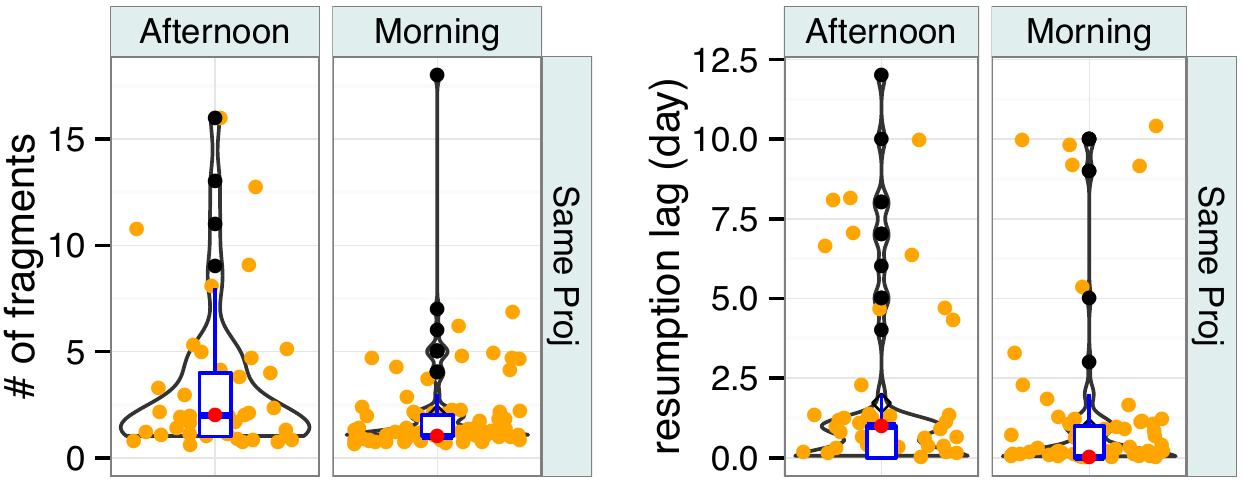}
        \vspace{-3.5mm}
     { \captionof{figure}{{[P1]} {\it{{\scriptsize{\big<{\it (same project, daytime, afternoon), \(D_{1-2}\)}}}}\big >}}
 \label{fig:Violin}}
 \vspace{-3mm}
    \end{minipage}\hfill
\end{tcolorbox}

Likewise, 10 (71\%) experienced participants agreed that interruptions with a different priority negatively impacts their productivity on the primary task [P9]. 15 (75\%) participants agreed that both context switching and priority difference negatively impacts their productivity [P5]; {\it ``depends on the task. It helps that it is the same app but if the tasks are from very different epics than it can have a negative effect''.} Regardless of experience, 20 (80\%) participants agreed about the negative impact of context switching [P3, P7]. The only difference between these two patterns is their disruptiveness factor ({\it D}). 3 (60\%) less-experienced participants agreed on the negative impact of afternoon interruptions: {\it``I am more productive in the morning when no one is at the office (i.e. minimal distractions with no meetings)''} [P10].\\

\begin {figure}
\centering
\vspace{-.5mm}
\includegraphics[scale=0.32]{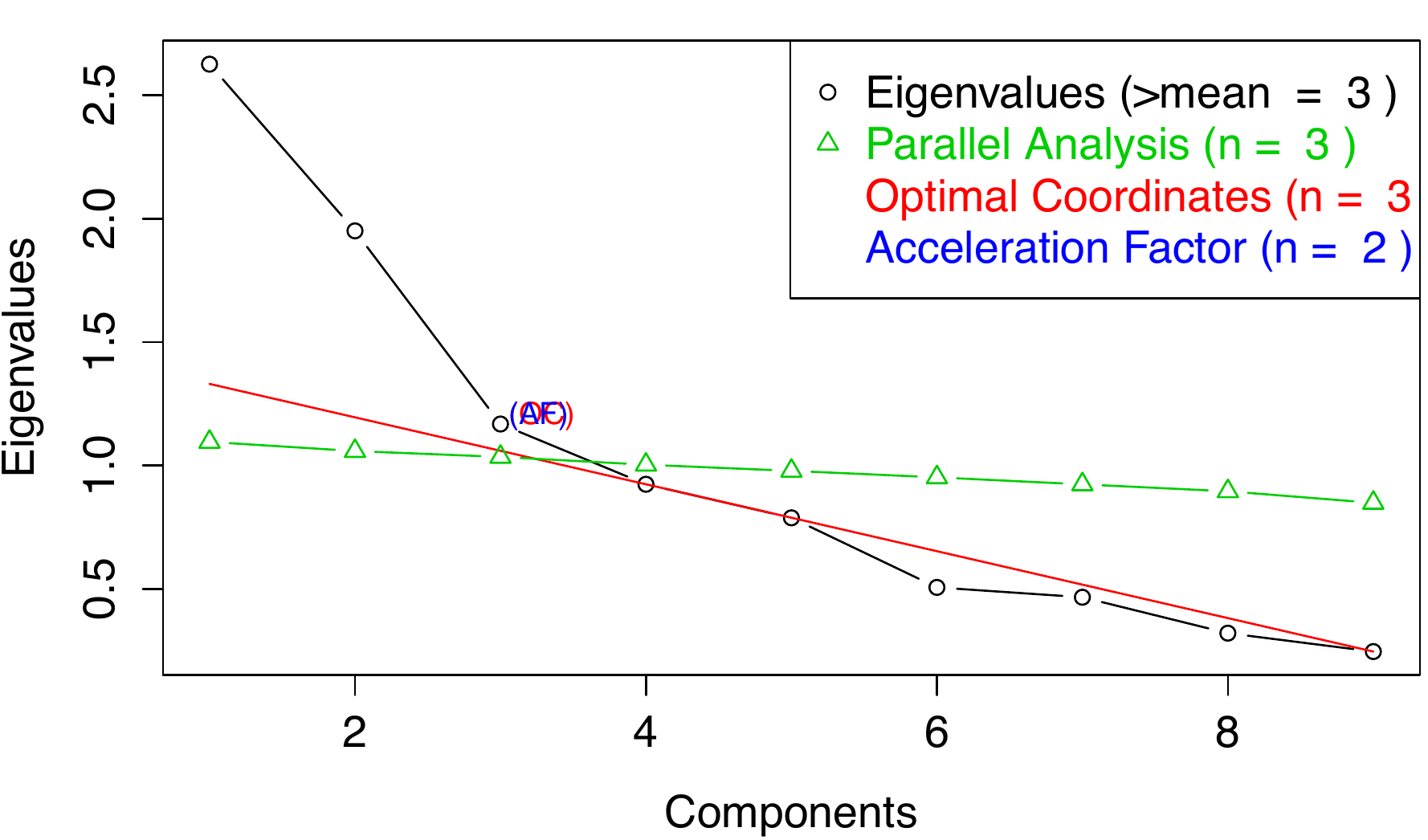}
\vspace{-3mm}
\caption{Number of Factors (Scree Test)}
\label{fig:scree}
\vspace{-7mm}
\end{figure}
\vspace{-3mm}
\section{Threats to Validity}
\label{sec:threats}

Conducting both studies on a particular company might question the generalizability of our findings. We attempted to address this threat by implementing the study on a fairly large dataset including various projects from different business domains and employees from different levels of experience. Moreover, applying a mixed method approach enabled us to triangulate findings obtained through the retrospective analysis with the results from the follow-up survey. Additionally, the employees' dataset we used in our retrospective analysis was recorded during their normal real-world work, not during an experimental exercise. 

Our data collection and preparation pose another threat to the validity of our results, since identifying the type of interruptions (i.e. self and external) and the type of the tasks in general (e.g. RE, architecture, development, test, ...) is not straightforward. To mitigate these risks, before the primary data collection phase, we pilot tested our data extraction method and the results of this stage have been validated by our industry partner. Moreover, the dataset associated with each employee was reviewed by at least two of the hired RA's and the first author of the paper. To evaluate the reliability of our decisions for independent variables that have been recorded manually, we used the CohenÕs Kappa statistic, which calculates the degree of agreement between two raters. The calculated Kappa
value was 0.96, which shows significant agreement, as stated by Landis and Koch \cite{kappa}.  

%Moreover, our data collection approach (i.e. analysis of work databases) strengthen the reliability and the internal validity of our study results \cite{DataC}.

Furthermore, the robustness and reliability of the measures (i.e. \(D_{1}, D_2\)) we used to study the disruptiveness of interruptions may affect the findings of this study. To define these underlying factors, we utilized the results of similar studies on interruption analysis (Section \ref{sec:terminology}). In addition, to ensure the reliability of these factors, we asked our survey participants about the negative impact of ``higher number of task fragments (\(D_1\))'' and ``longer resumption lags (\(D_2\))'' on their productivity. 23 (91\%) and 21 (85\%) participants agreed about the negative impact of both factors on their productivity after resuming the primary task. Further, to determine the number of factors we first used the {\it Scree test} (Figure \ref{fig:scree}) in conjunction with eigenvalues. Then, we manually checked the result of BEFA for the other number of factors (i.e. \(n= 2, 4, 5, 6\)). To ensure the reliability of the explored factors in our study and to minimize the impact of our data collection errors on these factors, we used a fairly large dataset (i.e. \(1,430\) interrupted tasks) for implementing the BEFA method, which is in the range of  {\it``\(\geq1000\) (excellent)''}, based on the sample size adequacy scale provided by Comerly and Lee \cite{SampleSize}.

%
%We are fairly confident that the assignment of events to working spheres is relatively accurate due to using multiple sources of information to assign them and validate them. 
%Our study has several limitations. First,
%
%Research Design is the process of selecting a method for a particular research problem, tapping into its strengths, while mitigating its weaknesses. The validity of the results depends on how well the research design compensates for the weaknesses of the methods \cite{RMethods}.
%
%The major weakness of case studies is that the data collection and analysis is more open to interpretation and researcher bias. For this reason, an explicit frame- work is needed for selecting cases and collecting data \cite{RMethods}.
%
%
%Another limitation is in the generalizability of the results. As we observed only one organization, we can only generalize our results to companies with similar characteristics: high pressure firms where many different tools are integral to work, and where people manage multiple activities. Further research is needed to understand work fragmentation in other work environments \cite{behind}.

%----------------------

 \section{Conclusion and Research Agenda}
 \label{sec:Conclusion}
 To investigate the concepts and the disruptiveness of task switching and interruptions in the area of RE, we conducted two studies, a manual retrospective analysis on 5,076 recorded tasks of 19 employees and a survey with 25 professional software developers. From both studies, we found that context switching, the priority of the interrupting task, the interruption source and timing are the key factors that negatively impact RE interruptions. We also provided a set of RE task switching patterns along with practical recommendations for practitioners in the area of RE and software development in general. 

Although our investigation provides new insights on RE task switchings and interruptions, our results raise additional research questions. Using visualization techniques as visual aids and tool support to control context switches and interruptions is one of these notable unanswered questions. For instance, {\it filling gaps} and {\it continuity} in the storytelling approach, as we discussed in our previous work \cite{SLR} (presented at RE 2016), increases developers' awareness of their transition between various tasks and may help developers in retrospection on their own productivity. We also asked our survey participants about the main features of a tool which would help them to have less disruptive interruptions. The majority of participants opined that the ability to record the next step of the primary task before switching to the new task should be the key feature of this tool, as in: {\it``It should allow me to make a quick note on what the next step was before the interruption so I could read it back and instantly know what the next step is''.} A stopwatch, timeline of a task, and possibly a historical view of the primary task are other features of a tool suggested by survey participants. 

As our future work, we aim to replicate this study in other contexts and to extend our investigation by looking closer at the impact on productivity and projects' success as dependent variables. In addition, we plan to further investigate the task type by applying requirements classification techniques (as in \cite{DataTrack}), and consider the task size as another independent variable, as suggested by our industry partner. Moreover, measuring the cognitive load of interruptions and proposing visualization techniques \cite{RENext17} to reduce the cognitive demands of task interruptions are other goals for replicating this study.

\section{Acknowledgement}
This research was supported by the Natural Sciences and
Engineering Research Council of Canada (NSERC) Discovery
Grant 250343-12  and by an Engage Grant. Also, we sincerely thank Arcurve Inc. for providing their comprehensive task management dataset and all Arcurve employees for their valuable input to our research and willingness to cooperate.

\bibliographystyle{IEEEtran}
\bibliography{IEEEabrv,RE17}

 \newcommand{\noop}[1]{}
\begin{thebibliography}{10}
\providecommand{\url}[1]{#1}
\csname url@samestyle\endcsname
\providecommand{\newblock}{\relax}
\providecommand{\bibinfo}[2]{#2}
\providecommand{\BIBentrySTDinterwordspacing}{\spaceskip=0pt\relax}
\providecommand{\BIBentryALTinterwordstretchfactor}{4}
\providecommand{\BIBentryALTinterwordspacing}{\spaceskip=\fontdimen2\font plus
\BIBentryALTinterwordstretchfactor\fontdimen3\font minus
  \fontdimen4\font\relax}
\providecommand{\BIBforeignlanguage}[2]{{%
\expandafter\ifx\csname l@#1\endcsname\relax
\typeout{** WARNING: IEEEtran.bst: No hyphenation pattern has been}%
\typeout{** loaded for the language `#1'. Using the pattern for}%
\typeout{** the default language instead.}%
\else
\language=\csname l@#1\endcsname
\fi
#2}}
\providecommand{\BIBdecl}{\relax}
\BIBdecl

\bibitem{Resumption}
C.~Parnin and S.~Rugaber, ``Resumption strategies for interrupted programming
  tasks,'' \emph{Software Quality Journal}, vol.~19, no.~1, pp. 5--34, 2011.

\bibitem{Mind}
D.~D. Salvucci and N.~A. Taatgen, \emph{The multitasking mind}.\hskip 1em plus
  0.5em minus 0.4em\relax Oxford University Press, 2010.

\bibitem{RO}
Z.~S.~H. Abad and G.~Ruhe, ``{Using Real Options to Manage Technical Debt in
  Requirements Engineering},'' in \emph{IEEE 23rd International Requirements
  Engineering Conference (RE'15)}, 2015, pp. 230--235.

\bibitem{Olli}
O.~Karras, S.~Kiesling, and K.~Schneider, ``{Supporting Requirements
  Elicitation by Tool-Supported Video Analysis},'' in \emph{IEEE 24th
  International Requirements Engineering Conference (RE)}, 2016, pp. 146--155.

\bibitem{Parisa}
P.~Ghazi and M.~Glinz, ``{An Exploratory Study on User Interaction Challenges
  When Handling Interconnected Requirements Artifacts of Various Sizes},'' in
  \emph{IEEE 24th International Requirements Engineering Conference (RE)},
  2016, pp. 76--85.

\bibitem{ParisaZahra}
P.~Ghazi, Z.~S.~H. Abad, and M.~Glinz, ``{Choosing Requirements for
  Experimentation with User Interfaces of Requirements Modeling Tools},'' in
  \emph{25th {IEEE} International Requirements Engineering Conference (RE'17)},
  2017.

\bibitem{memoryofgoals}
E.~M. Altmann and J.~Trafton, ``{Memory for Goals: An Activation-based
  Model},'' \emph{Cognitive Science}, vol.~26, no.~1, pp. 39 -- 83, 2002.

\bibitem{Disruptive2}
T.~Gillie and D.~Broadbent, ``What makes interruptions disruptive? a study of
  length, similarity, and complexity,'' \emph{Psychological Research}, vol.~50,
  no.~4, pp. 243--250, 1989.

\bibitem{ResumptionLag}
E.~M. Altmann and J.~G. Trafton, ``Task interruption: Resumption lag and the
  role of cues,'' DTIC Document, Tech. Rep., 2004.

\bibitem{Multitasking2}
D.~D. Salvucci and P.~Bogunovich, ``Multitasking and monotasking: The effects
  of mental workload on deferred task interruptions,'' in \emph{Proceedings of
  the SIGCHI Conference on Human Factors in Computing Systems}.\hskip 1em plus
  0.5em minus 0.4em\relax ACM, 2010, pp. 85--88.

\bibitem{DecisionMaking}
C.~Speier, I.~Vessey, and J.~S. Valacich, ``The effects of interruptions, task
  complexity, and information presentation on computer-supported
  decision-making performance,'' \emph{Decision Sciences}, vol.~34, no.~4, pp.
  771--797, 2003.

\bibitem{Concurrent}
D.~D. Salvucci, N.~A. Taatgen, and J.~P. Borst, ``Toward a unified theory of
  the multitasking continuum: From concurrent performance to task switching,
  interruption, and resumption,'' in \emph{Proceedings of the SIGCHI Conference
  on Human Factors in Computing Systems}, ser. CHI '09.\hskip 1em plus 0.5em
  minus 0.4em\relax ACM, 2009, pp. 1819--1828.

\bibitem{Relation}
E.~Cutrell, M.~Czerwinski, and E.~Horvitz, ``Notification, disruption, and
  memory: Effects of messaging interruptions on memory and performance.''\hskip
  1em plus 0.5em minus 0.4em\relax IOS Press, 2001, pp. 263--269.

\bibitem{Disruptive}
S.~M. Hess and M.~C. Detweiler, ``{Training to Reduce the Disruptive Effects of
  Interruptions},'' in \emph{Proceedings of the Human Factors and Ergonomics
  Society Annual Meeting}, vol.~38, no.~18.\hskip 1em plus 0.5em minus
  0.4em\relax SAGE Publications, 1994, pp. 1173--1177.

\bibitem{behind}
G.~Mark, V.~M. Gonzalez, and J.~Harris, ``No task left behind?: Examining the
  nature of fragmented work,'' in \emph{Proceedings of the SIGCHI Conference on
  Human Factors in Computing Systems}, ser. CHI '05.\hskip 1em plus 0.5em minus
  0.4em\relax ACM, 2005, pp. 321--330.

\bibitem{Gonzalez}
V.~M. Gonz\'{a}lez and G.~Mark, ``"constant, constant, multi-tasking
  craziness": Managing multiple working spheres,'' in \emph{Proceedings of the
  SIGCHI Conference on Human Factors in Computing Systems}, ser. CHI '04.\hskip
  1em plus 0.5em minus 0.4em\relax ACM, 2004, pp. 113--120.

\bibitem{selfinterruption}
I.~Katidioti, J.~P. Borst, and N.~A. Taatgen, ``What happens when we switch
  tasks: Pupil dilation in multitasking.'' \emph{Journal of experimental
  psychology: applied}, vol.~20, no.~4, p. 380, 2014.

\bibitem{Instant}
M.~Czerwinski, E.~Cutrell, and E.~Horvitz, ``Instant messaging: Effects of
  relevance and timing,'' in \emph{People and computers XIV: Proceedings of
  HCI}, vol.~2, 2000, pp. 71--76.

\bibitem{Temporal}
C.~A. Monk, D.~A. Boehm-Davis, and J.~G. Trafton, ``The attentional costs of
  interrupting task performance at various stages,'' in \emph{Proceedings of
  the Human Factors and Ergonomics Society Annual Meeting}, vol.~46,
  no.~22.\hskip 1em plus 0.5em minus 0.4em\relax SAGE Publications, 2002, pp.
  1824--1828.

\bibitem{ICSE}
B.~Vasilescu, K.~Blincoe, Q.~Xuan, C.~Casalnuovo, D.~Damian, P.~Devanbu, and
  V.~Filkov, ``{The Sky is Not the Limit: Multitasking Across GitHub
  Projects},'' in \emph{Proceedings of the 38th International Conference on
  Software Engineering}.\hskip 1em plus 0.5em minus 0.4em\relax ACM, 2016, pp.
  994--1005.

\bibitem{fragments}
H.~Sanchez, R.~Robbes, and V.~M. Gonzalez, ``An empirical study of work
  fragmentation in software evolution tasks,'' in \emph{IEEE 22nd International
  Conference on Software Analysis, Evolution, and Reengineering}, 2015, pp.
  251--260.

\bibitem{Zimmer}
A.~N. Meyer, T.~Fritz, G.~C. Murphy, and T.~Zimmermann, ``{Software Developers'
  Perceptions of Productivity},'' in \emph{Proceedings of the 22Nd ACM SIGSOFT
  International Symposium on Foundations of Software Engineering}, ser. FSE
  2014.\hskip 1em plus 0.5em minus 0.4em\relax ACM, 2014, pp. 19--29.

\bibitem{Pair}
J.~Chong and R.~Siino, ``{Interruptions on Software Teams: A Comparison of
  Paired and Solo Programmers},'' in \emph{Proceedings of the 2006 20th
  Anniversary Conference on Computer Supported Cooperative Work}.\hskip 1em
  plus 0.5em minus 0.4em\relax ACM, 2006, pp. 29--38.

\bibitem{Methodology}
P.~Runeson and M.~H{\"o}st, ``Guidelines for conducting and reporting case
  study research in software engineering,'' \emph{Empirical Software
  Engineering}, vol.~14, no.~2, pp. 131--164, 2008.

\bibitem{SERIP}
Z.~S.~H. Abad, G.~Ruhe, and M.~Bauer, ``{Understanding Task Interruptions in
  Service Oriented Software Development Projects: An Exploratory Study},'' in
  \emph{Proceedings of the 4th International Workshop on Software Engineering
  Research and Industrial Practice}, ser. SER\&IP '17.\hskip 1em plus 0.5em
  minus 0.4em\relax IEEE Press, 2017, pp. 34--40.

\bibitem{thumb}
B.~G. {Tabachnick et al}, ``Using multivariate statistics,'' 2001.

\bibitem{qq}
M.~B. Wilk and R.~Gnanadesikan, ``{Probability Plotting Methods for the
  Analysis of Data},'' \emph{Biometrika}, vol.~55, no.~1, pp. 1--17, 1968.

\bibitem{GT}
A.~Strauss and J.~M. Corbin, \emph{Grounded Theory in practice}.\hskip 1em plus
  0.5em minus 0.4em\relax Sage, 1997.

\bibitem{correlation}
D.~E. Hinkle, W.~Wiersma, and S.~G. Jurs, ``{Applied Statistics for the
  Behavioral Sciences},'' 2003.

\bibitem{kappa}
J.~R. Landis and G.~G. Koch, ``The measurement of observer agreement for
  categorical data,'' \emph{biometrics}, pp. 159--174, 1977.

\bibitem{SampleSize}
A.~L. Comrey and H.~B. Lee, \emph{A first course in factor analysis}.\hskip 1em
  plus 0.5em minus 0.4em\relax Psychology Press, 2013.

\bibitem{SLR}
Z.~S.~H. Abad, M.~Noaeen, and G.~Ruhe, ``{Requirements Engineering
  Visualization: A Systematic Literature Review},'' in \emph{2016 IEEE 24th
  International Requirements Engineering Conference (RE)}, 2016, pp. 6--15.

\bibitem{DataTrack}
Z.~S.~H. Abad, O.~Karras, P.~Ghazi, M.~Glinz, G.~Ruhe, and K.~Schneider,
  ``{What Works Better? A Study of Classifying Requirements},'' in
  \emph{Proceedings of the 25th IEEE International Conference on Requirements
  Engineering (RE'17)}, 2017.

\bibitem{RENext17}
Z.~S.~H. Abad, A.~Shymka, J.~Le, N.~Hammad, and G.~Ruhe, ``{A Visual Narrative
  Path from Switching to Resuming a Requirements Engineering Task},'' in
  \emph{2017 IEEE 25th International Requirements Engineering Conference
  (RE'17)}, 2017.

\end{thebibliography}

%\printbibliography
\end{document}